\documentclass[aps, pra,twocolumn,notitlepage, showpacs,superscriptaddress,10pt,floatfix]{revtex4-2}

\usepackage{graphicx}
\usepackage{grffile}
\usepackage{amsthm}
\usepackage{epsfig}
\usepackage{amsmath, amssymb, amsfonts, mathtools, bbm, MnSymbol, mathrsfs, xfrac}
\usepackage{longtable}
\usepackage{tabularx}
\usepackage{multirow}
\usepackage{makecell}
\usepackage{hhline}
\usepackage{colortbl}
\usepackage[table,svgnames,xcdraw]{xcolor}
\usepackage{array, graphicx}
\usepackage[breaklinks=true,colorlinks=true,linkcolor=blue,urlcolor=blue,citecolor=blue]{hyperref}
\usepackage{comment}
\usepackage{color}
\usepackage[makeroom]{cancel}

\usepackage{tikz}
\usetikzlibrary{decorations.pathreplacing,calligraphy}
\usepackage{pgfplots}
\usepackage{tikz-3dplot}
\tdplotsetmaincoords{60}{115}
\pgfplotsset{compat=newest}

\usepackage{soul}
\pdfoutput=1

\renewcommand{\[}{\begin{equation}}
\renewcommand{\]}{\end{equation}}
\renewcommand{\Re}{\mathfrak{Re}}
\renewcommand{\Im}{\mathfrak{Im}}
\newcommand{\ket}[1]{|#1\rangle}

\newcommand{\pro}[2]{|#1\rangle\!\langle#2|}
\newcommand{\mean}[1]{\left\langle#1\right\rangle}

\newcommand{\tr}{\mathrm{tr}}

\newcommand{\R}{{\hat{\rho}}}

\newcommand{\I}{{\hat{I}}}

\newcommand{\U}{{\hat{U}}}  

\newcommand{\floor}[1]{\left\lfloor #1 \right\rfloor}

\newcommand{\x}{\hat{x}}
\newcommand{\p}{\hat{p}}
\newcommand{\ad}{\mathrm{ad}}

\newcommand{\icol}[1]{\left(\begin{smallmatrix}#1\end{smallmatrix}\right)}

\newcommand{\irow}[1]{% inline row vector
  \begin{smallmatrix}(#1)\end{smallmatrix}%
}

\theoremstyle{definition}
\newtheorem{definition}{Definition}
\newtheorem{theorem}{Theorem}

\usepackage{physics}
\usepackage[page]{appendix}

\newcommand{\an}{\hat{a}}

\newcommand{\N}{\hat{N}}
\newcommand{\go}{{g\!\!\:o}}

\pdfoutput=1

\begin{document}

\title{
% Working title: Algebra of Gaussian operators and homomorphism with super conformal Poincaré algebra\\
% General Gaussian evolution and algebra, and homomorphism with super conformal Poincaré algebra\\
% Gaussian Quantum Dynamics and Their Algebra: A Homomorphism to Superconformal Symmetry\\
% General Gaussian Evolution and Its Algebraic Structure: A Homomorphism to the Superconformal Poincaré Algebra\\
% Gaussian Quantum Dynamics and Their Algebra: A Homomorphism to the Superconformal Poincaré Algebra\\
% Gaussian Quantum Dynamics and Its Algebraic Structure: A Homomorphism to the Superconformal Poincaré Algebra\\
% Gaussian Quantum Dynamics and Its Algebraic Structure\\
% Gaussian Quantum Dynamics and its Homomorphism to Superconformal Symmetry\\
% Gaussian quantum dynamics and its isomorphism with the Superconformal Poincaré Algebra\\
Gaussian Open Quantum Dynamics and Isomorphism to Superconformal Symmetry
}

\author{Ju-Yeon Gyhm}
\email{kjy3665@snu.ac.kr}
\affiliation{Department of Physics and Astronomy, Seoul National University, 1 Gwanak-ro, Seoul 08826, Korea}

\author{Dario Rosa}
\affiliation{ICTP South American Institute for Fundamental Research \\
Instituto de F\'{i}sica Te\'{o}rica, UNESP - Univ. Estadual Paulista \\
Rua Dr. Bento Teobaldo Ferraz 271, 01140-070, S\~{a}o Paulo, SP, Brazil}
\author{Dominik \v{S}afr\'{a}nek}
\affiliation{Center for Theoretical Physics of Complex Systems, Institute for Basic Science (IBS), Daejeon - 34126, Korea}

\date{\today}
\begin{abstract}
Understanding the Lie algebraic structure of a physical problem often makes it easier to find its solution. In this paper, we focus on the Lie algebra of Gaussian-conserving superoperators. We construct a Lie algebra of $n$-mode states, $\mathfrak{go}(n)$, composed of all superoperators conserving Gaussianity, and we find it isomorphic to $\mathbb{R}^{2n^2+3n}\oplus_{\mathrm{S}}\mathfrak{gl}(2n,\mathbb{R})$. This allows us to solve the quadratic-order Redfield equation for any, even non-Gaussian, state. We find that the algebraic structure of Gaussian operations is the same as that of super-Poincar\'e algebra in three-dimensional spacetime, where the CPTP condition corresponds to the combination of causality and directionality of time flow. Additionally, we find that a bosonic density matrix satisfies both the Klein-Gordon and the Dirac equations. Finally, we expand the algebra of Gaussian superoperators even further by relaxing the CPTP condition. We find that it is isomorphic to a superconformal algebra, which represents the maximal symmetry of the field theory. This suggests a deeper connection between two seemingly unrelated fields, with the potential to transform problems from one domain into another where they may be more easily solved.
\end{abstract}
\maketitle

\section{introduction}

Gaussian states are central to continuous-variable quantum information science and quantum optics due to their practical accessibility~\cite{Weedbrook2012, Olivares2012, Braunstein2005}. Coherent, squeezed, and thermal states, typical examples of Gaussian states, can be prepared and manipulated using standard optical components such as beam splitters, phase shifters, and squeezers~\cite{LEONHARDT199589}. This experimental feasibility makes them indispensable resources in a variety of quantum technologies, including quantum communication~\cite{Smith2011, DiGuglielmo2007}, Gaussian Boson sampling~\cite{Hamilton2017, ZHONG2019511, Quesada2018}, and continuous-variable quantum computing~\cite{Marshall2016, Gu2009, Lloyd1999}.

In addition to their experimental advantages, Gaussian states offer significant mathematical convenience. Their Wigner functions exhibit a Gaussian profile, and their full quantum description is captured entirely by the first and second statistical moments~\cite{Xu2016,Jiang2014}, greatly simplifying both analytical and numerical analysis~\cite{Graefe_2018}. Moreover, Gaussian states often exhibit properties that closely resemble classical behavior, such as positive Wigner functions and classical equations of  motion~\cite{Weedbrook2012, Olivares2012}. Due to these features, Gaussian states play a central role in quantum information processing. Accordingly, it is important to understand and develop open quantum dynamics that preserve Gaussianity~\cite{Eisert1,BRODIER20102315,Noh2019}. 

%In real-world settings, quantum systems inevitably interact with their environments, leading to decoherence and dissipation. Preserving the Gaussian character of states under such open-system dynamics is crucial for ensuring the reliability and efficiency of quantum protocols. Moreover, these dynamics offer a tractable and experimentally feasible framework for analyzing quantum evolutions in noisy environments.

Despite their significance, the theoretical characterization of Gaussian-preserving open quantum processes remains incomplete. Various approaches, such as third quantization~\cite{Prosen_2008,McDonald2023, Barthel_2022} and the pseudomode method~\cite{Pleasance2020, Tamascelli2018}, have been proposed to study open quantum systems. The third quantization provides a path integral formula for a Lindbladian, which can be challenging to use. On the other hand, pseduomode method can describe non-Markovian and both Gaussian and non-Gaussian evolution, but it does not fully capture how non-Gaussian states evolve under Gaussian-preserving dynamics.

Addressing this question with easy-to-use tools is particularly relevant because non-Gaussian states are indispensable resources in advanced quantum technologies. This includes bosonic quantum error correction~\cite{Gottesman2001, Grimsmo2020,Chamberland2022,Hastrup2022, Michael2016}, universal quantum computing~\cite{Bartlett2002, Menicucci2006, Walschaers2021,Ghose15042007}, quantum metrology~\cite{Huang2018, Dowling01032008}, and quantum batteries~\cite{Gian2024}, which rely on their non-Gaussian features to surpass classical limits.

In this work, we address this challenge by first relaxing the standard complete positivity and trace-preservation (CPTP) condition, which allows us to identify the Lie algebra underlying Gaussian-preserving open quantum dynamics. This algebraic perspective enables a more general class of dynamical generators that preserve Gaussianity, providing a rigorous framework for analyzing the evolution of non-Gaussian states in open systems. Then we reintroduce the CPTP condition to focus on quantum channels.

Using this framework, we derive general solutions to both the quadratic order Redfield and Lindblad equations, even for initial non-Gaussian states.

Finally, we uncover a connection with quantum field theory. Specifically, we show that the algebra governing the Gaussian-preserving single-mode dynamics is isomorphic to the $\mathcal{N}=1$ superconformal algebra in three-dimensional Minkowski spacetime. Moreover, both the Klein–Gordon and Dirac equations arise in this representation. Furthermore, we find that the CPTP condition is mathematically equivalent to the requirement of causal propagation in the three-dimensional spacetime. These findings open a new direction in connecting open quantum dynamics and field theory.

In Section~\ref{section:Lie_algebra}, we review some concepts in Lie theory.
Section~\ref{section:single} defines the Gaussian open algebra for a single bosonic mode and investigates its general properties, which we then utilize to derive solutions to the quadratic Redfield equation.
In Section~\ref{section:multi}, we extend these results to the multi-mode scenario.
In Section~\ref{section:open_superconformal}, we establish the isomorphism between the Gaussian open algebra for a single mode and the super-Poincar\'e algebra, as well as explore other related structures. Finally, in Section~\ref{section:superconformal_algebra} we define the extended Gaussian open algebra isomorphic to the superconformal algebra.

\section{Preliminaries: concepts in Lie theory}\label{section:Lie_algebra}

Symmetries (either continuous or discrete) play a fundamental role in solving the equations of motion of a given dynamical system or in finding its eigenstates. In physics, continuous symmetries are mathematically described by Lie groups, which are therefore instrumental in shedding light on the dynamics of a physical system under investigation. Utilizing the Lie theory, it is possible to study a Lie group via its associated (and simpler) Lie algebra. Before diving into a Lie theory characterization of open Gaussian dynamics we provide an overview of some relevant concepts~\cite{Humphreys1972} to make the paper more self-contained.

First, let us recall the mathematical definition of a Lie algebra. 
\begin{definition}
    A Lie algebra is defined as a vector space $\mathfrak{g}$ over a field, $\mathbb{F}$, equipped with a binary operator, $[\bullet,\bullet]:\mathfrak{g}\times\mathfrak{g}\rightarrow\mathfrak{g}$  called the Lie bracket, that satisfies the following three conditions.
    \begin{enumerate}
        \item \emph{Bilinearity}: The Lie bracket satisfies, 
        \[\begin{split}
        [ag_1 +bg_2,g_3] &=a[g_1,g_3]+ b[g_2,g_3] \\
        [g_1, ag_2+bg_3] &=a[g_1,g_2]+ b[g_1,g_3],
        \end{split}\]
        for all $g_1,g_2,g_3 \in \mathfrak{g}$, and scalars $a,b\in \mathbb{F}$.
        \item \emph{Anticommutativity}: The Lie bracket satisfies, 
        \[
        [g_1,g_2] = -[g_2,g_1],
        \]
        for all $g_1,g_2 \in \mathfrak{g}$,
        \item \emph{The Jacobi identity}: The Lie bracket satisfies
        \[
        [g_1,[g_2,g_3]]+[g_2,[g_3,g_1]]+[g_3,[g_1,g_2]]=0,
        \]
        for all $g_1,g_2,g_3 \in \mathfrak{g}$
    \end{enumerate}
\end{definition}
In this work, we take the Lie bracket to be the commutator between quantum operators, \textit{i.e.} $[g_1,g_2] = g_1g_2-g_2g_1$, commonly used to describe quantum systems. It satisfies bilinearity, anticommutativity, and the Jacobi identity, and thus it forms a Lie bracket. 

A special case of a Lie algebra is an \emph{Abelian Lie algebra}. In an Abelian Lie algebra, $\mathfrak{g}$, the commutator, $[g_1,g_2]$, of any two elements, $g_1,g_2 \in \mathfrak{g}$, is always zero. Simply, we can write $[\mathfrak{g},\mathfrak{g}]=0$. For example, pure translations on flat space form an Abelian Lie algebra since their generators commute with each other. An $ n$-dimensional Abelian Lie algebra over $\mathbb{R}$ will be denoted by $\mathbb{R}^n$. 

Next, we also introduce the concepts of Lie subalgebras and ideals in Lie algebras. Let us consider a linear subspace $\mathfrak{h} \subset \mathfrak{g}$. Lie subalgebras and ideals are defined as follows,
\begin{definition}
    $\mathfrak{h}\subset \mathfrak{g}$ is a \emph{Lie subalgebra} of $\mathfrak{g}$, if $[h_1,h_2] \in \mathfrak{h}$ for all $h_1,h_2\in \mathfrak{h}$.
\end{definition}
\begin{definition}
    $\mathfrak{h}\subset \mathfrak{g}$ is an \emph{ideal} in $\mathfrak{g}$, if $[h,g] \in \mathfrak{h}$ for all $h\in \mathfrak{h}$ and $g\in \mathfrak{g}$.
\end{definition}
These two concepts are useful for exploring the global structure of a Lie algebra, $\mathfrak{g}$. A familiar illustrative example, the Lie algebra of the Euclidean group, can provide a clearer understanding of these two concepts.
Consider the three-dimensional Euclidean group, $\mathrm{EU}(3)$, whose Lie algebra is denoted $\mathfrak{eu}(3)$. Within $\mathfrak{eu}(3)$, the subspace corresponding to pure rotations is precisely the special orthogonal Lie algebra $\mathfrak{so}(3)\subset \mathfrak{eu}(3)$. To verify that $\mathfrak{so}(3)$ is indeed a Lie subalgebra, observe that for all $h_1,h_2 \in \mathfrak{so}(3)$, the commutator, $[h_1,h_2]$ is again an element of $\mathfrak{so}(3)$. Therefore,  $\mathfrak{so}(3)$ is closed under commutation and constitutes a Lie subalgebra of $\mathfrak{eu}(3)$.

Furthermore, $\mathfrak{eu}(3)$ contains a subspace corresponding to pure translations, namely $\mathbb{R}^3\subset \mathfrak{eu}(3)$. Pure translations are closed under both rotations and other translations; equivalently, for all $h\in\mathbb{R}^3$, and $g\in \mathfrak{eu}(3)$, the commutator $[h,g]$ lies in $\mathbb{R}^3$; Hence $\mathbb{R}^3$ is an ideal in $\mathfrak{eu}(3)$

The Lie algebra $\mathfrak{eu}(3)$ also serves as an example for illustrating Lie algebra extensions. The definition of Lie algebra extensions is as follows,
\begin{definition}
    We say that
    $\mathfrak{e}=\mathfrak{h}\oplus_\mathrm{S}\mathfrak{g}$ is a Lie algebra extension of $\mathfrak{g}$ by $\mathfrak{h}$, (a semi direct sum of $\mathfrak{h}$ and $\mathfrak{g}$),
    if the vector space $\mathfrak{e} = \mathfrak{h}\times\mathfrak{g}$ satisfies the following properties: 
    \begin{enumerate}
    \item $\mathfrak{e}$ is a Lie algebra.
    \item $\mathfrak{h}$ is an ideal in $\mathfrak{e}$.
    \item $\mathfrak{g}$ is a Lie subalgebra of $\mathfrak{e}$.
    \end{enumerate}
\end{definition}
We have already shown that the translation subspace, $\mathbb{R}^3$, is an ideal in $\mathfrak{eu}(3)$ and that the rotation algebra, $\mathfrak{so}(3)$, is a Lie subalgebra of $\mathfrak{eu}(3)$. Moreover, 
$\mathfrak{eu}(3)$ can be decomposed into $\mathfrak{so}(3)$ and $\mathbb{R}^3$. By definition, $\mathfrak{eu}(3)$ is a Lie algebra extension of $\mathfrak{so}(3)$ by $\mathbb{R}^3$, which is written as $\mathfrak{eu}(3) = \mathbb{R}^3\oplus_\mathrm{S}\mathfrak{so}(3)$. 
The Euclidean group appears as a symmetry in many dynamical and static systems, which suggests that we are already making use of the concept of Lie algebra extensions.

Finally, we turn to the notion of \textit{Lie algebra isomorphism}, which plays a crucial role in classifying Lie algebras
\begin{definition}
    A Lie algebra $\mathfrak{g}$ is said to be isomorphic to another Lie algebra $\mathfrak{h}$ as $\mathfrak{g}\cong\mathfrak{h}$, if there exists a bijective linear map $f:\mathfrak{g}\rightarrow \mathfrak{h}$ that preserves the commutation relations. That is, for all $g_1,g_2\in\mathfrak{g}$, $f$ satisfies 
    \[
    [f(g_1),f(g_2)]=f([g_1,g_2]) \in \mathfrak{h}.
    \]
    Under this condition, $f$ is called a Lie algebra isomorphism.
\end{definition}
It is well known that $\mathfrak{so}(3)$ is isomorphic to $\mathfrak{su}(2)$, which is often used in quantum mechanics~\cite{WEYL1966}. Lie algebra isomorphisms enable us to construct more efficient representations of a given Lie algebra.
    
\section{Gaussian open quantum dynamics of a single bosonic mode}\label{section:single}

Unitary Gaussian operators for a single bosonic mode are composed of the rotation operator, $\hat{R}(\theta)$, the squeezing operator, $\hat{S}(z)$, and the displacement operator, $\hat{D}(\alpha)$~\cite{Weedbrook2012}. All three are unitary operators, each generated by a corresponding Hamiltonian. The Hamiltonian generating the displacement operator is a linear combination of the position and momentum operators, which are themselves given in terms of the annihilation, $\an$, and creation $\an^\dagger$ as
\[
\hat{x}=\frac{1}{\sqrt{2}}(\an + \an^\dagger),\quad \hat{p}=\frac{i}{\sqrt{2}}(\an^\dagger-\an).
\]
The Hamiltonian corresponding to the rotation operator is the number operator,
\[
\label{eq:quadratic_Lie_elements_1}
\hat{N}=\tfrac{1}{2}(\an^{\dagger}\an+\an\an^\dagger),
\]
while the squeezing operator is generated by quadratic combinations of creation and annihilation operators,
\[
\label{eq:quadratic_Lie_elements_2}
\hat{X}=\tfrac{i}{2}(\an^{\dagger 2}-\an^2),\quad
\hat{Y}=\tfrac{1}{2}(\an^{\dagger 2}+\an^2).
\]

The set of Hamiltonians corresponding to unitary Gaussian operators is closed under commutation, up to constants. The relevant commutation relations are,
\[\begin{split}
[\hat{N},\hat{X}] = 2i\hat{Y}\quad [\hat{N},\hat{Y}] =- 2i\hat{X}\quad[\hat{X},\hat{Y}]=-2i\hat{N},
\end{split}
\]
and,
\[
[\x,\p]=i\hat{I}.
\]
The vector space spanned by $i\hat{N}$, $i\hat{X}$, and $i\hat{Y}$ forms a Lie algebra, which is isomorphic to the real symplectic Lie algebra~\cite{Braunstein2005},
\[
\mathrm{Span}(i\hat{N},i\hat{X},i\hat{Y})  \cong\mathfrak{sp}(2,\mathbb{R})
\]
Meanwhile, the vector space spanned by $i\x$, $i\p$, and $i$  also forms a Lie algebra
\[
\mathfrak{w} = \mathrm{Span}(i\x,i\p,i\hat{I}) 
\]
which corresponds to the Weyl–Heisenberg algebra~\cite{Weyl1927, vonNeumann2018}.

The vector space spanned by all Hamiltonians associated with unitary Gaussian operators is given by~\cite{Braunstein2005}
\[\label{eq:Gaussian_unitary_generators}
\mathfrak{ug}\equiv\mathrm{Span}(i\hat{N},i\hat{X},i\hat{Y},i\x,i\p,i\hat{I}) \cong \mathfrak{w} \times\mathfrak{sp}(2,\mathbb{R})
\]
One can confirm that $\mathfrak{w}$ is an ideal in $\mathfrak{ug}$ and $\mathfrak{sp}(2,\mathbb{R})$ is a Lie subalgebra of $\mathfrak{ug}$. As a result, $\mathfrak{ug}$ forms a Lie algebra extension, which can be expressed as, 
\[
\mathfrak{ug} = \mathfrak{w}\oplus_\mathrm{S}\mathfrak{sp}(2,\mathbb{R}).
\]

$\mathfrak{ug}$ is the maximal set of Hamiltonians that preserve the Gaussianity of quantum states~\cite{Weedbrook2012}. Moreover, the algebraic structure of $\mathfrak{ug}$ has also been extensively studied, even in the case of multi-mode bosonic systems~\cite{Braunstein2005}. However, we point out that there exist additional operators corresponding to open quantum dynamics that also preserve Gaussianity. Moreover, they also satisfy closed commutation relations among themselves, forming a Lie algebra.

To describe the open quantum dynamics, we need to propose the concept of superoperators. For a closed quantum system, we consider operators acting on pure states $\ket{\psi}$, which are elements of the Hilbert space, $\mathcal{H}$. In contrast, open quantum systems are described using operators that act on the density matrix, $\R$, where $\R$ is an element of the space of bounded operators on $\mathcal{H}$, denoted $\mathcal{B}(\mathcal{H})$. Therefore, to describe open dynamics in this setting, we require operators that act on operators; these are known as \emph{superoperators}~\cite{KRAUS1971311, Sudarshan1961, Nielsen_Chuang_2010}.

The dynamics of the density matrix  $\R$ is governed by the Lindbladian superoperator~\cite{Lindblad1976}, $\mathcal{L}$, expressed as
\[\begin{split}
\frac{d}{dt}\R = \mathcal{L}\R =i[\R,\hat{H}]+\sum_j D(\hat{A_j})\R,
\end{split}
\]
where the term $i[\R,\hat{H}]$ corresponds to the unitary part of the evolution, and $D(\hat{A}_j)$ represents the dissipative part, defined by
\[
D(\hat{A}_j)\R=\hat{A}_j\R \hat{A}_j^\dagger - \tfrac{1}{2}\{\hat{A}_j^\dagger \hat{A}_j, \R\}.
\]
Note that the Hamiltonian $\hat{H}$ itself is not a superoperator. Instead, the superoperator associated with unitary evolution is given by
\[
\ad_{\hat{H}} = i [\hat{H},\,\bullet\,],
\] 
which is referred to as the adjoint operator of $\hat{H}$.
This adjoint superoperator generates the unitary conjugation as $\exp(\ad_{\hat{H}})=\hat{U} \bullet \hat{U}^\dagger$, where $\hat{U}=\exp(i\hat{H})$. For instance, the adjoint operators corresponding to Hamiltonians in $\mathfrak{ug}$ generate the unitary evolution of the density matrix, as follows
\begin{align}
        \hat{R}(\theta)\R\hat{R}^\dagger(\theta)&=\exp\left(-\theta\ad_{\hat{N}}\right)\R,\\
    \hat{S}(z)\R\hat{S}^\dagger(z)&=\exp\left(\Re(z) \ad_{\hat{X}}-\Im(z)\ad_{\hat{Y}}\right)\R,\\
    \hat{D}(\alpha)\R\hat{D}^\dagger(\alpha)&=\exp(\sqrt{2}(\Im(\alpha)\ad_{\hat{x}}-\Re(\alpha) \ad_{\p}))\R,
\end{align}

where these expressions represent the action of unitary Gaussian operators on the density matrix, $\R$. As a result, $\ad_{\x}$, $\ad_{\p}$, $\ad_{\hat{N}}$, $\ad_{\hat{X}}$, and $\ad_{\hat{Y}}$ are the superoperators corresponding to the unitary Gaussian operators.

Other Lindbladian superoperators that preserve Gaussianity include one-photon noise processes. We define the following superoperators, 
\begin{eqnarray}\label{eq:quadratic_Lie_elements2}
    L_{\hat{O}_1 \hat{O}_2}^+ &=&\hat{O}_1\,\bullet\,\hat{O}_2 + \hat{O}_2\,\bullet\,\hat{O}_1 -\tfrac{1}{2}\{\{\hat{O}_2,\hat{O}_1\},\,\bullet\, \}\\
    L_{\hat{O}_1 \hat{O}_2}^- &=&i\hat{O}_1\,\bullet\,\hat{O}_2 -i \hat{O}_2\,\bullet\,\hat{O}_1 -\tfrac{1}{2}\{i[\hat{O}_2,\hat{O}_1],\,\bullet\, \}.
\end{eqnarray}
The following four superoperators correspond to various types of one-photon dissipation,
\begin{eqnarray}
    L_{\x\x}^+&=&2\x\bullet\x -\{\x^2,\bullet\},\\
    L_{\p\p}^+&=&2\p\bullet\x -\{\p^2,\bullet\},\\
    L_{\x\p}^+&=&\x\bullet\p+\p\bullet\x -\tfrac{1}{2}\{\x\p+\p\x,\bullet\},\\
    L_{\x\p}^-&=&i\x\bullet\p-i\p\bullet\x -\tfrac{1}{2}\{\hat{I},\bullet\}.
\end{eqnarray}
The conventional Lindblad master equation can be expressed as a linear combination of these superoperators. We find, for example, that the zero-temperature thermalization equation~\cite{Breuer2002} is given by
\[
\frac{d}{dt}\R=\gamma\big(\an\R\an^\dagger +\tfrac{1}{2}\{\an^\dagger\an, \R\}\big)=\gamma\big(\tfrac{1}{4}L_{\x\x}^++\tfrac{1}{4}L_{\p\p}^+-\tfrac{1}{2}L_{\x\p}^{-}\big)\R.
\]

The superoperators $L_{\hat{x}\hat{x}}^+$, $L_{\hat{x}\hat{p}}^+$, and $L_{\hat{p}\hat{p}}^+$ with positive parity can be associated with diffusion processes induced by real-valued classical noise. These terms appear naturally in quantum Fokker–Planck equations~\cite{Breuer2002}. In contrast, the operator $L_{\hat{x}\hat{p}}^-$ exhibits odd parity and does not correspond to a standard diffusion process. Its interpretation requires more care. One heuristic view is to associate it with an effective noise term whose statistical correlations are purely imaginary, drawing an analogy to analytic continuations of classical Brownian motion. While not physical in a strict stochastic sense, such terms do arise in formal treatments of non-Hermitian or non-equilibrium open quantum systems. 

Alternatively, and more concretely, $L_{\hat{x}\hat{p}}^-$ can be interpreted as generating a scaling (or dilation) transformation in phase space. When acting on the Wigner function, it induces a uniform rescaling of the phase-space coordinates, modifying the width of the distribution while preserving its overall shape. This interpretation aligns with the symplectic structure of the phase-space formulation and highlights the role of $L_{\hat{x}\hat{p}}^-$ in non-diffusive, non-Hamiltonian dynamics.

We define the vector space, $\mathfrak{go}(1)$, spanned by the Gaussianity preserving superoperators, as follows.
\begin{definition}
    $\mathfrak{go}(1)$ is vector space defined as
    \[\label{eq:def_go(1)}
    \mathfrak{go}(1)\!\equiv\!\mathrm{Span}(\!\ad_{\x}, \ad_{\p}, \ad_{\hat{N}},\ad_{\hat{X}},\ad_{\hat{Y}},L_{\x\x}^+ ,L_{\p\p}^+ ,L_{\x\p}^+ ,L_{\x\p}^- ),
    \]
    for a single bosonic mode.
\end{definition}
We refer to $\mathfrak{go}(1)$ as the \emph{Gaussian open algebra}~\footnote{We choose the name Gaussian open algebra to signify that this algebra is defined in terms of superoperators, which also describe open quantum dynamics. At the same time, the closed quantum dynamics is also covered, since it can also be rephrased in terms of superoperators.}. It satisfies the following theorem.
\begin{theorem}\label{theorem:go_is_Lie}
$\mathfrak{go}(1)$ is a Lie algebra. Further, it is isomorphic to $\mathbb{R}^5\oplus_{\mathrm{S}}\mathfrak{gl}(2,\mathbb{R})$. Each element of this algebra can be expressed as a linear combination,
\[\label{eq:most_general_Lie_element}
\begin{split}
\go\ &\equiv\ \gamma_{\hat{x}}\ad_{\hat{x}}+\gamma_{\hat{p}}\ad_{\hat{p}}+\gamma_{\hat{N}}\ad_{\hat{N}}+\gamma_{\hat{X}}\ad_{\hat{X}}+\gamma_{\hat{Y}}\ad_{\hat{Y}}\\
&+\ \gamma_{\x\x}^{+}L_{\x\x}^{+} +\gamma_{\x\p}^{+}L_{\x\p}^{+}+\gamma_{\p\p}^{+}L_{\p\p}^{+}+ \gamma_{\x\p}^{-}L_{\x\p}^{-},
\end{split}
\]
where $\gamma$'s are nine real numbers.
\end{theorem}
\begin{proof} (sketch, see Appendix~\ref{subsection:proof_of_iso_go(1)} for details)

We define
\[\begin{split}
    \mathfrak{g}&\equiv\mathrm{Span}(\ad_{\hat{N}},\ad_{\hat{X}},\ad_{\hat{Y}},L_{\x\p}^-)\\
    \mathfrak{h}&\equiv\mathrm{Span}(\ad_{\x},\ad_{\p},L_{\x\x}^+,L_{\p\p}^+,L_{\x\p}^+).
\end{split}
\]
We will show that $\mathfrak{g}\cong\mathfrak{gl}(2,\mathbb{R})$ and $\mathfrak{h}\cong \mathbb{R}^5$. It is obvious $\mathfrak{go}(1)=\mathfrak{h}\times\mathfrak{g}$. We find $[\mathfrak{h},\mathfrak{h}]=0$, which means that $\mathfrak{h}$ is a five-dimensional Abelian Lie algebra, $\mathbb{R}^5$. What remains is to determine the commutation relations $[\mathfrak{g},\mathfrak{g}]$ and $[\mathfrak{g},\mathfrak{h}]$.

To derive these commutation relations, we first express each of the superoperators in Eq.~\eqref{eq:def_go(1)} in terms of $\ad_{\x}$, $\ad_{\p}$, $\ad^+_{\x}$, and $\ad^+_{\p}$. $\ad_{\hat{O}}^+$ is the anti‐adjoint operator of an operator, $\hat{O}$, defined by
\[
\ad_{\hat{O}}^+ \R = \{\R,\hat{O}\} = \R\hat{O}+\hat{O}\R.
\]

First, doing this we find that every element $g\in\mathfrak{g}$ can be represented by a corresponding two-by-two real matrix, $\Gamma$, 
as 
\[
\begin{pmatrix}
-\ad_{\p} &
\ad_{\x} 
\end{pmatrix}
\Gamma\begin{pmatrix}
\frac{\ad^+_{\x}}{2} \\ \frac{\ad_{\p}^+}{2} 
\end{pmatrix} \in \mathfrak{g}.
\]
We find that 
\[\begin{split}
[g_1,g_2] =\begin{pmatrix}
-\ad_{\p} &
\ad_{\x} 
\end{pmatrix}[\Gamma_1,\Gamma_2]
\begin{pmatrix}
\frac{\ad^+_{\x}}{2} \\ \frac{\ad_{\p}^+}{2} 
\end{pmatrix}
\in \mathfrak{g}.
\end{split}
\]
Thus, the mapping $\mathfrak{g}\rightarrow \Gamma$ is a Lie algebra isomorphism and $\mathfrak{g}$ is a Lie algebra.

Second, each element $h_1\in\mathrm{Span}(\ad_{\x},\ad_{\p})\equiv\mathfrak{h}_1$, is represented a two dimensional vector, $\bold{v}$, as
\[\label{eq:ad_vector}
\begin{pmatrix}
    -\ad_{\p} & \ad_{\x} 
\end{pmatrix}\bold{v}=h_1\in \mathfrak{h}_1,
\]

With this, we derive the commutation relation between $g\in\mathfrak{g}$ and $h_1\in\mathfrak{h}_1$,
\[\begin{split}\label{eq:[g,h1]}
[g,h_1] =\begin{pmatrix}
    -\ad_{\p} & \ad_{\x} 
\end{pmatrix}(\Gamma\bold{v})\in \mathfrak{h}_1,
\end{split}
\]
Thus, $\mathfrak{h}_1$ is an ideal in $\mathfrak{go}(1)$.

Third, element $h_2\in\mathrm{Span}(L_{\x\x}^{+}, L_{\p\p}^{+}, L_{\x\p}^{+})\equiv\mathfrak{h}_2$ is represented by a two-by-two symmetric real matrix $D$, as
\[
 \begin{pmatrix}
-\ad_{\p} & \ad_{\x} \\
\end{pmatrix}D\begin{pmatrix}
-\ad_{\p} \\
\hphantom{-}\ad_{\x} \\
\end{pmatrix}=h_2\in\mathfrak{h}_2,
\]
We find the commutation relation between $h_2$ and $g\in\mathfrak{g}$,
\[\begin{split}\label{eq:[g,h2]_main}
[g,h_2] =\begin{pmatrix}
-\ad_{\p} & \ad_{\x} \\
\end{pmatrix}(\Gamma D+D\Gamma^\top)\begin{pmatrix}
-\ad_{\p} \\
\hphantom{-}\ad_{\x} \\
\end{pmatrix}\in \mathfrak{h}_2.
\end{split}
\]
Thus, $\mathfrak{h}_2$ is an ideal in $\mathfrak{go}(1)$. Since both  $\mathfrak{h}_1$ and $\mathfrak{h}_2$ are ideals,  $\mathfrak{h}=\mathfrak{h}_1\times\mathfrak{h}_2$ is also an ideal.

To summarize the results, $\mathfrak{g}\cong\mathfrak{gl}(2,\mathbb{R})$ is a Lie subalgebra of $\mathfrak{go}(1)$ and $\mathfrak{h}\cong \mathbb{R}^5$ is ideal in $\mathfrak{go}(1)$. These results imply that $\mathfrak{go}(1)$ is isomorphic to an extended Lie algebra, $\mathfrak{gl}(2,\mathbb{R})$ by $\mathbb{R}^5$, denoted $\mathbb{R}^5\oplus_\mathrm{S}\mathfrak{gl}(2,\mathbb{R})$.
\end{proof}

Theorem~\ref{theorem:go_is_Lie} shows that a commutator of every two elements from $\mathfrak{go}(1)$ also falls into $\mathfrak{go}(1)$. Moreover, it provides the isomorphism between $\mathfrak{go}(1)$ and extended general Lie algebra, $\mathfrak{gl}(2)$, by a five dimensional abelian Lie algebra, $\mathbb{R}^5$.
We organize the structure of $\mathfrak{go}(1)$ in Table~\ref{table:GGO}.

\begin{table}[]
\begin{tabular}{ccccc}
\hline
\multicolumn{5}{|c|}{Gaussian superoperators}                                                   \\ \hhline{|=====|}
\multicolumn{3}{|c|}{Gaussian operator}  & \multicolumn{2}{c|}{One photon noise}               \\ 
\multicolumn{3}{|c|}{(Unitary operator)} & \multicolumn{2}{c|}{(non-Unitary operator)}   \\ \hline
\multicolumn{1}{|c|}{Rotation} & \multicolumn{1}{c|}{Squeezing} & \multicolumn{1}{c|}{Displacement} & \multicolumn{1}{c|}{Real} & \multicolumn{1}{c|}{Imaginary} \\ \hline
\multicolumn{1}{|c|}{$\ad_{\hat{N}}$} & \multicolumn{1}{c|}{$\ad_{\hat{X}}$, $\ad_{\hat{Y}}$} & \multicolumn{1}{c|}{$\ad_{\x}$, $\ad_{\p}$} & \multicolumn{1}{c|}{$L_{\x\x}^{+}$, $L_{\p\p}^{+}$} & \multicolumn{1}{c|}{$L_{\x\p}^{-}$} \\
\multicolumn{1}{|c|}{} & \multicolumn{1}{c|}{} & \multicolumn{1}{c|}{} & \multicolumn{1}{c|}{$L_{\x\p}^{+}$} & \multicolumn{1}{c|}{}
\\\hhline{|=====|}
\multicolumn{5}{|c|}{Algebraic structure}                                                                                                         \\ \hline
\multicolumn{2}{|c|}{$\mathfrak{sp}(2,\mathbb{R})$} &\multicolumn{1}{c|}{$\mathbb{R}^2$} & \multicolumn{1}{c|}{$\mathbb{R}^3$}  & \multicolumn{1}{c|}{$\mathbb{R}$}                         \\ \hline
\multicolumn{3}{|c|}{ $\mathbb{R}^2\oplus_{\mathrm{S}}\mathfrak{sp}(2,\mathbb{R}) $} &\multicolumn{2}{c|}{$\mathbb{R}^3\oplus_{\mathrm{S}} \mathbb{R}$}                       \\ \hline
\multicolumn{5}{|c|}{ $\mathfrak{go}(1)\cong\mathbb{R}^5\oplus_{\mathrm{S}}\mathfrak{gl}(2,\mathbb{R}) $}                      \\ \hline
\end{tabular}
\caption{The table about the basis of $\mathfrak{go}(1)$ and their algebraic structure.}
\label{table:GGO}
\end{table}

Although $\mathfrak{go}(1)$ is a well-defined Lie algebra, not every element of $\mathfrak{go}(1)$ corresponds to a valid Lindbladian superoperator. Indeed, the Lindbladian superoperator must satisfy the CPTP condition. We provide the following theorem to confirm the CPTP condition for the elements of $\mathfrak{go}(1)$
\begin{theorem}\label{theorem:cptp}(CPTP condition)
The superoperator $\go\in\mathfrak{go}(1)$ is always trace-preserving (TP). Additionally, it is completely positive (CP) if and only if these two conditions are satisfied:
\begin{subequations}
\begin{align}
4\gamma_{\x\x}^+\gamma_{\p\p}^+&\geq (\gamma_{xp}^{+})^2+(\gamma_{xp}^{-})^2,\\
\gamma_{\x\x}^++\gamma_{\p\p}^+&\geq 0.
\end{align}
\end{subequations}
\end{theorem}
\begin{proof}
    See Appendix~\ref{subsection:proof_cptp}.
\end{proof}
The theorem allows us to determine whether a given element of $\mathfrak{go}(1)$ corresponds to a Lindbladian superoperator.

It follows from the relationship between Lie groups and Lie algebras that the evolution generated by an element \(\go \in \mathfrak{go}(1)\) can be expressed as a (not necessarily CP) superoperator:
\begin{equation}
\mathcal{E}_G = \exp(\go) \in \mathrm{GO}^+(1),
\end{equation}
where we define the Lie group
\begin{equation}\label{eq:go+1isomorphism}
\mathrm{GO}^+(1) \equiv \left\{ \exp(\go) \;\middle|\; \go \in \mathfrak{go}(1) \right\}\cong \mathbb{R}^5\oplus_{\mathrm{S}}\mathrm{GL}^{+}(2,\mathbb{R}),
\end{equation}
which corresponds to the Lie algebra \(\mathfrak{go}(1)\). $\mathrm{GL}^{+}(2,\mathbb{R})$ is a real matrix group with a strictly positive determinant. Notably, \(\mathfrak{go}(1)\) is nine-dimensional, meaning that only nine real parameters are needed to characterize \(\mathcal{E}_G\), even though it acts on an infinite-dimensional Hilbert space. This parameter efficiency is a central feature that underpins the Gaussian evolution of quantum states.

Although \(\mathrm{GO}^+(1)\) is a well-defined Lie group and related to Gaussian states, it is not complete. Specifically, while \(\mathrm{GO}^+(1)\) can connect most pairs of Gaussian states, it cannot transform a mixed state into a pure state, although it can asymptotically approach such a transformation, e.g., in the open system dynamics at infinite times. This incompleteness arises from the fact that \(\mathrm{GO}^+(1)\) is an open set and therefore does not include all limit points of its elements.

To resolve this limitation, we introduce the \emph{topological closure} of \(\mathrm{GO}^+(1)\), denoted by \(\overline{\mathrm{GO}^+(1)}\). The topological closure is the smallest closed set containing \(\mathrm{GO}^+(1)\), and it includes all limits of converging sequences from \(\mathrm{GO}^+(1)\). Formally, we define~\footnote{It is important to note that $\overline{\mathrm{GO}^+(1)}$ is not equal to $\mathbb{R}^5\oplus_{\mathrm{S}}\overline{\mathrm{GL}^+(2,\mathbb{R})}$, While $\overline{\mathrm{GL}^+(2,\mathbb{R})}$ denotes the topological closure of the general linear group 
$\mathrm{GL}^+(2,\mathbb{R})$ within the topology of real matrices. $\overline{\mathrm{GO}^+(1)}$ refers to the closure of a group of superoperators acting on density matrices, and must therefore be understood within the topology of superoperators. $\overline{\mathrm{GO}^+(1)}$ should be defined as a topological closure in the strong operator topology, not merely in the matrix space~\cite{Holevo2019, Shirokov2020, Shirokov2021}.
},
\begin{equation}
\overline{\mathrm{GO}^+(1)} \equiv \mathrm{GO}^+(1) \cup \left\{ \lim_{k \to \infty} \mathcal{E}_{G,k} \;\middle|\; \forall k \in \mathbb{N},\; \mathcal{E}_{G,k} \in \mathrm{GO}^+(1) \right\}.
\end{equation}

This completion \(\overline{\mathrm{GO}^+(1)}\) forms the full set of (possibly limiting) Gaussian evolutions and provides a rigorous mathematical framework to describe transformations that \(\mathrm{GO}^+(1)\) alone cannot capture~\footnote{The following theorem represents a form of the third thermodynamic law: without the closure, it is not possible to create a pure state from mixed by a finite number of operations.}. The introduction of \(\overline{\mathrm{GO}^+(1)}\) leads to the following theorem.

\begin{theorem}\label{theorem:existence_of_GGO}
For any two Gaussian states, $\R$ and $\R'$, there exists $\mathcal{E}_G\in \overline{\mathrm{GO}^+(1)}$ satisfying the CPTP condition, which connects the states as $\R' =\mathcal{E}_G\R$.
\end{theorem}
\begin{proof}
    See Appendix~\ref{section:proof_existence}.
\end{proof}

Note that an element of $\mathfrak{go}(1)$ satisfying CPTP corresponds to a Lindblad superoperator, and that an element of $\overline{\mathrm{GO}^+(1)}$ satisfying CPTP corresponds to the exponential of such a Lindblad superoperator or its limit. Combining this fact with the result of Theorem~\ref{theorem:existence_of_GGO}, we conclude that there exists a Lindbladian evolution, $\mathcal{E}(t)=e^{\mathcal{L}t}$, that connects any pair of Gaussian states.

The fact that $\overline{\mathrm{GO}^+(1)}$ is complete in the space of Gaussian states naturally leads us to the concept of a \emph{Gaussian quantum channel}. Gaussian quantum channels are defined as linear superoperators that (a) are CPTP maps, and (b) transform every Gaussian state into another Gaussian state. We now investigate the relationship between $\overline{\mathrm{GO}^+(1)}$ and Gaussian quantum channels.

\begin{theorem}\label{proposition:Gaussian_channel}
The following statements are equivalent:
\begin{enumerate}
    \item $\mathcal{E}$ is an infinitesimally divisible Gaussian quantum channel~\cite{Teiko2009}.
    \item $\mathcal{E}$ is an element of $\overline{\mathrm{GO}^+(1)}$ and satisfies the CP condition.
\end{enumerate}    
\end{theorem}
\begin{proof}
    See Appendix~\ref{section:proof_of_GO+}.
\end{proof}

Here, \emph{infinitesimally divisible} refers to a quantum channel that can be expressed as the exponential of a Lindbladian superoperator, or as the limit of such exponentials~\cite{Teiko2009}.

As a consequence, we find that $\overline{\mathrm{GO}^+(1)}$ encompasses most Gaussian quantum channels, particularly those that are infinitesimally divisible. However, there still exist Gaussian quantum channels that lie outside $\overline{\mathrm{GO}^+(1)}$.

This limitation arises from a fundamental property of Lie theory: the exponential map from a Lie algebra generates only a connected component of the identity in the corresponding Lie group. Therefore, to fully represent the group of Gaussian quantum channels, additional generators beyond those in the original Lie algebra are required.

We identify the \emph{transpose channel}, $\mathcal{E}_{\top}$, as one such additional generator. It is defined by
\[
\mathcal{E}_{\top}(\hat{\rho}) = \hat{\rho}^{\top}
\]
for all density matrices $\hat{\rho}$. The transpose channel preserves the Gaussianity of quantum states. Although the transpose channel is positive and trace-preserving but not a CPTP map, it plays a crucial role in extending the structure to encompass the full set of Gaussian quantum channels.

Analogously to the definition of $\mathrm{GO}^+(1)$, we define the conjugate Lie group,
\[
\mathrm{GO}^-(1) \equiv \left\{ \exp(\go) \, \mathcal{E}_{\top} \,\middle|\, \go \in \mathfrak{go}(1) \right\}\cong\mathbb{R}^5\oplus_{\mathrm{S}}\mathrm{GL}^{-}(2,\mathbb{R}).
\]
The full group $\mathrm{GO}(1)$ is then defined as the union:
\[
\mathrm{GO}(1) = \mathrm{GO}^+(1) \cup \mathrm{GO}^- (1) \cong\mathbb{R}^5\oplus_{\mathrm{S}}\mathrm{GL}(2,\mathbb{R}).
\]
Furthermore, we define the topological closure $\overline{\mathrm{GO}(1)}$, which we show to fully cover the space of Gaussian quantum channels.

\begin{theorem}\label{proposition:Gaussian_channel_2}
The following statements are equivalent:
\begin{enumerate}
    \item $\mathcal{E}$ is a Gaussian quantum channel.
    \item $\mathcal{E}$ is an element of $\overline{\mathrm{GO}(1)}$ and satisfies the CP condition.
\end{enumerate}    
\end{theorem}
\begin{proof}
    See Appendix~\ref{section:proof_Gassian=GO}.
\end{proof}

While $\overline{\mathrm{GO}^+(1)}$ captures all infinitesimally divisible Gaussian channels, $\overline{\mathrm{GO}(1)}$
is required to describe the full space of Gaussian quantum channels, including those that are not connected to the identity component.
As a consequence, we conclude that the set of Gaussian quantum channels is a strict subset of $\overline{\mathrm{GO}(1)}$. This is because Gaussian quantum channels form a semigroup, like general quantum channels, lacking inverse elements. In contrast, $\overline{\mathrm{GO}(1)}$ contains the full Lie group~\footnote{The inverse element of $\exp(\go)$ is $\exp(-\go)$, but it is larger than the Lie group, because some elements on the boundary do not have an inverse element.}.

As an example of application of the formalism just developed, we find a general solution for the time-dependent Redfield equation~\cite{Breuer2002} with at most quadratic order in the creation and annihilation operators,

\[\label{eq:most_general_eom_GGO}
\begin{split}
    \frac{d\R}{dt} &= \mathcal{R}\R = i[h_{\an^2}\an^2+h_{\an^\dagger\an}\an^\dagger\an+h_{\an}\an+\mathrm{h.c.},\R]\\
    &\!\!\!\!\!+\gamma_{\an\an^\dagger}(\an\R\an^\dagger-\tfrac{1}{2}\{\an^\dagger\an, \R\})+\gamma_{\an^\dagger \an}(\an\R\an^\dagger-\tfrac{1}{2}\{\an^\dagger\an, \R\})\\
    &\!\!\!\!\!+\gamma_{\an\an}(\an\R\an-\tfrac{1}{2}\{\an^2, \R\})+\gamma_{\an\an}^*(\an^\dagger\R\an^\dagger-\tfrac{1}{2}\{\an^{\dagger 2}, \R\}).
\end{split}
\]
The coefficients $h$ and $\gamma$ can be time-dependent. $h_{\an^2}$, $h_{\an}$, and $\gamma_{\an\an}$ are complex while the other coefficients are real. The Redfield equation may not satisfy the CPTP condition, however, the Redfield superoperator, $\mathcal{R}$, is an element of $\mathfrak{go}(1)$. 
As a result, we consider a general evolution of type
\[\label{eq:general_evolution}
\begin{split}
    \frac{d\R}{dt}=\mathcal{R}\R=\go(t)\R.
    % \go(t)&\equiv\gamma_{\hat{N}}(t)\ad_{\hat{N}}+\cdots+\gamma_{\x\x}^{+}(t)L_{\x\x}^{+} +\cdots+ \gamma_{\x\p}^{-}(t)L_{\x\p}^{-}
\end{split}
\]
When the Redfield superoperator meets the CPTP condition, it can be rewritten as the Lindbladian superoperator satisfying the CPTP condition given by Theorem~\ref{theorem:cptp}.

The evolution has a known solution as a time-ordered exponential,
\[
\R(t)=\mathcal{T}\exp \left(\int_0^t \go(\tau)d\tau \right)\R(0) = \mathcal{E}_G(t) \R(0).
\]
It follows from the Lie theory that the evolution operator $\mathcal{E}_G(t)$ also belongs to the Lie group $\mathrm{GO}(1)$. The dimension of both $\mathcal{E}_G$ and $\R$ is infinite in the Fock space representation. This makes this expression difficult to work with. Utilizing the isomorphism given by Eq.~\eqref{eq:go+1isomorphism} allows to us represent $\mathcal{E}_G(t)$ with finite matrices~\footnote{For unitary Gaussian operators, $\mathcal{E}_G(t)\cong \mathrm{Sp}(2,\mathbb{R})\oplus_{\mathrm{S}}\mathbb{R}^2$, which matches with well known result about the Gaussian evolution under the closed dynamics, which are described by the symplectic group (for the squeezing and rotations) together with the displacement, in the phase-space formalism of Gaussian states~\cite{Weedbrook2012,Adesso2010,Dominik2016}}.

$\mathcal{E}_G(t)$ can be represented by a tuple $(M,D,\bold{v})$, where $M\cong \mathrm{GL}^+(2,\mathbb{R})$ is a real matrix, $D\cong \mathbb{R}^3$ is a real symmetric matrix, and $\bold{v}\cong \mathbb{R}^2$ is a real vector. The identity superoperator, which represents $\mathcal{E}_G(0)$, is given by
\[\label{eq:identity_rep}
M=\begin{pmatrix}
    1 & 0\\
    0 & 1
\end{pmatrix},\quad D=\begin{pmatrix}
    0 & 0\\
    0 & 0
\end{pmatrix},\quad \bold{v}=\begin{pmatrix}
    0 \\
    0 
\end{pmatrix}.
\]

Defining
\[
\Gamma(t)=\begin{pmatrix}
\gamma_{\x\p}^{-}-\gamma_{\hat{X}} &  \gamma_{\hat{Y}} -\gamma_{\hat{N}} \\
\gamma_{\hat{Y}}+\gamma_{\hat{N}}  & \gamma_{\x\p}^{-}+\gamma_{\hat{X}} 
\end{pmatrix},
\]
the time evolution of $\mathcal{E}_G(t)=(D(t)\oplus\bold{v}(t))\oplus_{\mathrm{S}} M(t)$ is 
\[\label{eq:evolution_redfield}
\begin{split}
    \frac{d}{dt}M&= \Gamma M,\\
    \frac{d}{dt}{D}&=\begin{pmatrix}
    \gamma_{\p\p}^{+}  & -\gamma_{\x\p}^{+}/2 \\
    -\gamma_{\x\p}^{+}/2  & \gamma_{\x\x}^{+} 
\end{pmatrix}  +\Gamma D+D\Gamma^\top,\\
    \frac{d}{dt}{\bold{v}}&=\begin{pmatrix}
    -\gamma_{\p}\\
    \hphantom{-}\gamma_{\x}
    \end{pmatrix}+\Gamma\bold{v}.
\end{split}
\]

With this, we can write a general solution for the dynamics of any, even non-Gaussian, initial state $\R(0)$.
\begin{theorem}\label{theorem:general_sol_single_mode}
    For any initial quantum state $\R(0)$ described by the Wigner function $W(x,p)_0$, the general solution for the dynamics $\mathcal{E}_G=(D\oplus\bold{v})\oplus_{\mathrm{S}} M$ expressed in Eq.~\eqref{eq:general_evolution} is given by
\[\label{eq:Wigner_transform}\begin{split}
    W(x,p)_t\!=&\frac{1}{|\mathrm{det}(M)|}\!\int_{-\infty}^\infty \!\!\int_{-\infty}^\infty  dx' dp' W\left(M^{-1}\left(\icol{
    x'\\
    p'}-\bold{v}\right)\right)_0\\
    &\times\frac{1}{4\pi\sqrt{\mathrm{det}(D)}}\exp({-\irow{x-x'&p-p'}\tfrac{D^{-1}}{4}\icol{x-x'\\p-p'}}).
\end{split}
\]
\end{theorem}
\begin{proof}
    See Appendix~\ref{section:evolution_GO}.
\end{proof}

This theorem also describes transformation of the Wigner function under a general $\overline{\mathrm{GO}(1)}$ element~\footnote{This is done by considering the parametrization~\eqref{eq:most_general_Lie_element} and defining time-evolved channel $\mathcal{E}_G(t)=\exp(\go\,  t)$, meaning that every coefficient is transformed $\gamma \rightarrow \gamma t$. Then we solve Eqs.~\eqref{eq:evolution_redfield}, transform the Wigner function according to Theorem~\ref{theorem:general_sol_single_mode}, and set $t=1$.}.

%One can notice that the Wigner only comprises linear coordinate change and free diffusion. 

We make three observations. First, this type of dynamics transforms the Wigner function solely through a linear change of coordinates and free diffusion as can be seen from the change in the arguments. Second, while the Gaussian evolution in the closed dynamics is represented by the symplectic structure, for the open dynamics this structure breaks down. Unitary Gaussian operators have Lie algebra $\mathbb{R}^2\oplus\mathfrak{sp}(2,\mathbb{R})$ which is extended to $\mathbb{R}^5\oplus\mathfrak{gl}(2,\mathbb{R})$ for non-unitary operators. 
Third, given that it is a general solution, it can also be applied to any non-Gaussian state such as cat states, Gottesman-Kitaev-Preskill (GKP) state, and the number state with non-trivial integer, $N$, which are considered valuable resources in quantum technologies~\cite{Gottesman2001, Gian2024, Deng2024, Schlegel_2022, Hastrup2022}.

\section{Multi bosonic mode}\label{section:multi}

Next, we generalize the Gaussian open algebra to multi-mode systems, $\mathfrak{go}(n)$. Consider an $n$-mode bosonic system. For each mode $i=1,\cdots,n$, one defines a local Gaussian open algebra, $\mathfrak{go}(1)_i$, obtained from $\mathfrak{go}(1)$ by replacing the canonical operators, $\x$, and $\p$, with their mode indexed counterparts $\x_i$, and $\p_i$. In addition, for every pair of distinct modes, $i<j$, the bi-mode algebra, $\mathfrak{go}(2)_{ij}$ is generated by the adjoint operator, $\ad_{\hat{N}_{ij+}}$, $\ad_{\hat{N}_{ij-}}$, corresponding to beam splitting (mode-mixing) operators, and  $\ad_{\hat{X}_{ij}}$, $\ad_{\hat{Y}_{ji}}$, corresponding to multi-mode squeezing operators. The corresponding Hamiltonians can be expressed in terms of creation and annihilation operators as,
\[\begin{split}\label{eq:bi-mode_unitary}
    \hat{N}_{ij+}&=\tfrac{1}{2}(\an^\dagger_i\an_j+\an^\dagger_j\an_i),\quad
    \hat{N}_{ij-}=\tfrac{i}{2}(\an^\dagger_i\an_j-\an^\dagger_j\an_i),\\
    \hat{X}_{ij}&=\tfrac{i}{2}(\an^\dagger_i\an^\dagger_j-\an_i\an_j),\quad
    \ \ \hat{Y}_{ij}=\tfrac{1}{2}(\an^\dagger_i\an^\dagger_j+\an_i\an_j).
\end{split}
\]

$\mathfrak{go}(2)_{ij}$ also contains the dissipative one-photon superoperators,
 \[\begin{split}\label{eq:bi-mode_noise}
L_{\x_i\x_j}^+,\quad L_{\x_i\p_j}^+,\quad L_{\x_j\p_i}^+, \quad L_{\p_i\p_j}^+,\\
L_{\x_i\x_j}^-,\quad L_{\x_i\p_j}^-,\quad L_{\x_j\p_i}^-, \quad L_{\p_i\p_j}^-,
\end{split}
\]
which generalize the real and imaginary noise Brownian terms to bi‐mode couplings. Hence, $\mathfrak{go}(2)_{ij}$ is defined by twelve-dimensional vector space as
\[
\mathfrak{go}(2)_{ij}=\mathrm{Span}(\ad_{\hat{N}_{ij+}},\cdots, L_{\x_i\x_j}^+, \cdots, L_{\p_i\p_j}^- ),
\]
which consist of the superoperators in Eq.~\eqref{eq:bi-mode_unitary} and Eq.~\eqref{eq:bi-mode_noise}. These basis vectors are organized in Table~\ref{table:bi-mode_GGO}.

 The full $n$-mode Gaussian open algebra is defined by
\[
\mathfrak{go}(n)=\bigoplus_{1\leq i\leq n} \mathfrak{go}(1)_i \oplus\bigoplus_{1\leq i<j\leq n} \mathfrak{go}(2)_{ij}.
\]
As in the single‐mode case, $\mathfrak{go}(n)$ is a well‐defined Lie algebra.
\begin{theorem}\label{theorem:lie_go(n)}
    $\mathfrak{go}(n)$ is a Lie algebra isomorphic to $\mathbb{R}^{2n^2+3n}\oplus_{\mathrm{S}}\mathfrak{gl}(2n,\mathbb{R}) $. 
\end{theorem}
\begin{proof} (sketch)
We define
\[\begin{split}
\mathfrak{g}&\equiv\mathrm{Span}(\ad_{\hat{N_i}}\!,\ad_{\hat{X_i}}\!,\ad_{\hat{Y_i}}\!,\ad_{\hat{N}_{ij+}},\ad_{\hat{N}_{ij-}},\ad_{\hat{X}_{ij}},\\
&\ \ \ \ \ \ \ \ \ \ \ \ad_{\hat{Y}_{ji}}, L_{\x_i\p_j}^-,L_{\x_i\x_j}^-,L_{\p_i\p_j}^-),\\
\mathfrak{h}&\equiv\mathrm{Span}(\ad_{\x_i},\ad_{\p_i},L_{\x_i\x_j}^+,L_{\p_i\p_j}^+,L_{\x_i\p_j}^+).
\end{split}
\]
We show that $\mathfrak{g}\cong\mathfrak{gl}(2n,\mathbb{R})$ and $\mathfrak{h}\cong \mathbb{R}^{2n^2+3n}$. See Appendix~\ref{subsection:proof_of_go(n)}. 
\end{proof}

Furthermore, 
Theorem~\ref{theorem:existence_of_GGO},
Theorem~\ref{proposition:Gaussian_channel}, Theorem~\ref{proposition:Gaussian_channel_2}, 
and Theorem~\ref{theorem:general_sol_single_mode} all admit straightforward generalizations to $\mathfrak{go}(n)$. In the physical aspect, Theorem~\ref{theorem:existence_of_GGO} and Theorem~\ref{proposition:Gaussian_channel_2} imply that any open‐system dynamics that preserves Gaussianity, regardless of the number of modes, must be generated by elements of $\mathfrak{go}(n)$.

In particular, the generalization of Theorem~\ref{theorem:general_sol_single_mode} to the multi‐mode setting shows that open dynamics generated by $\mathfrak{go}(n)$ admit an efficient representation. Whenever a master equation is expressed by $\go(t)\in\mathfrak{go}(n)$, the time evolution superoperator can be written as
\[\label{eq:time_evolution_multimode}
\mathcal{E}_G(t)=\mathcal{T}\exp(\int_0^t\go(\tau) d\tau )\in \mathrm{GO}^+(n),
\]
Similar to a single mode, also for an $n$-mode bosonic system, any element in $\mathrm{GO}^+(n)$ is fully determined by $6n^2+3n$ real parameters. Hence, the description of open Gaussian dynamics scales \emph{quadratically} rather than exponentially in the number of modes.

Theorem~\ref{theorem:lie_go(n)} shows that any $\mathcal{E}_G\in \mathrm{GO}^+(n)$ admits a semi direct sum parametrization $\mathcal{E}_G\cong(D\oplus\bold{v})\oplus_{\mathrm{S}} M$, where $M\in\mathbb{R}^{2n\times2n}$, $D\in\mathbb{R}^{2n\times2n}$ is symmetric, and $\bold{v}$ is a real vector in $\mathbb{R}^{2n}$. To integrate the corresponding equation of motion, one must track the time evolution of $(D\oplus\bold{v})\oplus_{\mathrm{S}} M$ under a generator $\go(t)\in\mathfrak{go}(n)$.

To simplify the notation, we assemble the adjoint and anti‐adjoint operators into $2n$ component column vectors
\[\begin{split}
    \vec{\ad}=\begin{pmatrix}
    -\ad_{\p_1}\\
    \vdots\\
    -\ad_{\p_n}\\
    \hphantom{-}\ad_{\x_1}\\
    \vdots\\
    \hphantom{-}\ad_{\x_n}
\end{pmatrix},\quad
\vec{\ad^+}=\begin{pmatrix}
    \ad_{\x_1}^+\\
    \vdots\\
    \ad_{\x_n}^+\\
    \ad_{\p_1}^+\\
    \vdots\\
    \ad_{\p_n}^+
\end{pmatrix},\quad
\end{split}
\]
Any time‐dependent generator $\go(t)\in\mathfrak{go}(n)$ can be decomposed as,
\[
\go(t)=\frac{1}{2}
\vec{\ad}^\top \Gamma_M \vec{\ad^+}+\vec{\ad}^\top\Gamma_D \vec{\ad}+\vec{\ad}^\top
\Gamma_{\bold{v}},
\]
where $\Gamma_M\in \mathbb{R}^{2n\times2n}$, $\Gamma_D\in \mathbb{R}^{2n\times2n}$ is symmetric, and $\Gamma_\bold{v}\in \mathbb{R}^{2n}$. This block‐matrix representation makes the dynamics of $\mathcal{E}_G=(D\oplus \bold{v})\oplus_{\mathrm{S}} M$ possible to integrate. Indeed, from the original equation of the motion, $\frac{d}{dt}\mathcal{E}_G = \go(t)\mathcal{E}_G$ one obtains the decoupled equations for the parameters,
\[\begin{split}
    \frac{d}{dt}M &=  \Gamma_M M,\\
    \frac{d}{dt}D &= \Gamma_D + \Gamma_M D+D\Gamma_M^\top,\\
    \frac{d}{dt}\bold{v}&=\Gamma_{\bold{v}}+\Gamma_M\bold{v}.
\end{split}
\]
As in the single‐mode case, any Gaussian open evolution $\mathcal{E}_G$ is fully specified by the $M$, $D$, and $\bold{v}$. The solution for the dynamics follows.

\begin{theorem}\label{theorem:general_sol_multi}
    For any initial quantum state $\R(0)$ described by the Wigner function $W(\vec{x},\vec{p})_0$, the general solution for the dynamics $\mathcal{E}_G=(D\oplus\bold{v})\oplus_{\mathrm{S}} M$ is given by
    \[\begin{split}
    W(\vec{x},\vec{p})_t=\frac{1}{|\mathrm{det}(M)|}\int_{-\infty}^\infty  d\vec{y}^n d\vec{q}^n W\left(M^{-1}
    \left(\icol{\vec{y}\\\vec{q}}-\bold{v}\right)
    \right)_0\\
    \times\frac{1}{(4\pi)^n \sqrt{\mathrm{det}(D)}}e^{-\irow{\vec{x}^\top-\vec{y}^\top,\ \vec{p}^\top-\vec{q}^\top}\tfrac{D^{-1}}{4}\icol{\vec{x}-\vec{y}\\\vec{p}-\vec{q}}}.
\end{split}
\]
\end{theorem}

This expression makes explicit that the Gaussian‐open dynamics preserves the Gaussian form of $W(\vec{x},\vec{p})$, deforming it by a linear coordinate transformation followed by $M$, $\bold{v}$, and convolution with a Gaussian function determined by $D$.

\begin{table}[]
\begin{tabular}{cccc}
\hline
\multicolumn{4}{|c|}{Bi-mode superoperator acting on $i$,$j$ site}                                                   \\ \hhline{|====|}
\multicolumn{2}{|c|}{Gaussian operator}  & \multicolumn{2}{c|}{One photon noise}               \\ 

\multicolumn{2}{|c|}{(Unitary operator)} & \multicolumn{2}{c|}{(non-Unitary operator)}   \\ \hline
\multicolumn{1}{|c|}{Beam } &\multicolumn{1}{c|}{multi mode} & \multicolumn{1}{c|}{Real}& \multicolumn{1}{c|}{Imaginary}   \\
\multicolumn{1}{|c|}{splitter} &\multicolumn{1}{c|}{squeezing} & \multicolumn{1}{c|}{}& \multicolumn{1}{c|}{}   \\ \hline
\multicolumn{1}{|c|}{$\ad_{\hat{N}_{ij+}},$}&\multicolumn{1}{c|}{ $\ad_{\hat{X}_{ij}}$,} & \multicolumn{1}{c|}{$L_{\x_i\x_j}^+$, $L_{\p_i\p_j}^+$,}& \multicolumn{1}{c|}{$L_{\x_i\x_j}^-$, $L_{\p_i\p_j}^-$,}\\
\multicolumn{1}{|c|}{ $\ad_{\hat{N}_{ij-}}$} & \multicolumn{1}{c|}{ $\ad_{\hat{Y}_{ij}}$} & \multicolumn{1}{c|}{ $L_{\x_j\p_i}^+$, $L_{\x_i\p_j}^+$}& \multicolumn{1}{c|}{ $L_{\x_j\p_i}^-$, $L_{\x_i\p_j}^-$} \\ \hline
\end{tabular}\label{table:GGO_2}
\caption{The basis vectors of bi-mode Gaussian open algebra, $\mathfrak{go}(2)_{ij}$.}\label{table:bi-mode_GGO}
\end{table}

\section{duality between Gaussian open dynamics and three-dimensional field theory}\label{section:open_superconformal}

Lie algebras are often used to describe symmetries, a key ingredient in simplifying physical problems. In this section, we demonstrate that the Lie algebra we have constructed, $\mathfrak{go}(1)$, is isomorphic to three fundamental symmetries of modern physics: translational symmetry, Lorentz symmetry, and supersymmetry.

Given the Lie algebra $\mathfrak{go}(1) \cong \bold{R}^{5}\oplus_\mathrm{S} \mathfrak{gl}(2,\mathbb{R})$, we note that $\mathbb{R}^{3}\oplus_\mathrm{S} \mathfrak{sl}(2,\mathbb{R})$ forms a Lie subalgebra. Moreover, by recalling the isomorphism $\mathfrak{sl}(2,\mathbb{R})\cong\mathfrak{so}(2,1)$, we therefore conclude that the Lie algebra $\mathbb{R}^3\oplus_{\mathrm{S}}\mathfrak{so}(2,1)$, known as the three-dimensional \emph{Poincar\'e algebra}, forms a Lie subalgebra of $\mathfrak{go}(1)$.

Let us recall that the Poincar\'e algebra consists of the Lorentz operators in three-dimensional spacetime and the translation operators, so it can be written as
\[
\mathbb{R}^3\oplus_{\mathrm{S}}\mathfrak{so}(2,1) = \mathrm{Span}(T_{xy},T_{x\tau},T_{y\tau},P_\tau,P_x,P_y),
\]
where $T_{\mu\nu}$ are the Lorentz operators and $P_\mu$ are the momentum operators (which are the generators of translations). We denote the two spatial coordinates by $x$ and $y$ and the time coordinate by $\tau$. Moreover, since Einstein notation is used in this section, the distinction between upper and lower indices must be carefully observed, where we use the 
$(-,+,+)$ metric signature.

The Lie algebra isomorphism between  the Poincar\'e algebra and the $\mathfrak{go}(1)$ subalgebra $\mathbb{R}^{3}\oplus_\mathrm{S} \mathfrak{sl}(2,\mathbb{R})$ is realized at the level of generators by the following identifications 
\[\label{eq:duality}
\begin{split}
    P_\tau&\Leftrightarrow\tfrac{1}{2}(L_{\x\x}^+ +L_{\p\p}^+)\\
    P_x&\Leftrightarrow\tfrac{1}{2}(L_{\x\x}^+-L_{\p\p}^+)\\
    P_y&\Leftrightarrow  L_{\x\p}^+\\
    T_{xy}&\Leftrightarrow -\tfrac{1}{2}\ad_{\hat{N}}\\
    T_{\tau x}&\Leftrightarrow \tfrac{1}{2}\ad_{\hat{X}}\\
    T_{\tau y}&\Leftrightarrow -\tfrac{1}{2}\ad_{\hat{Y}}\, ,
\end{split}
\]
that preserve all the commutation relations. This bijective map therefore represents an isomorphism between the two algebras.

In this Section, we aim to explore the physical meaning of this isomorphism.
Let us consider a bosonic field in three dimensions, denoted by $\phi(x,y,\tau)$. We translate the field as
\begin{multline}\label{eq:translation_on_field}
\exp(\Delta x P_x+\Delta y P_y+\Delta \tau P_\tau)\phi(x,y,\tau) \\
= \phi(x+\Delta x,y+\Delta y,\tau+\Delta \tau) \, .
\end{multline}

Applying the identification in Eq.~\eqref{eq:duality}, we see that the translation operation of Eq.~\eqref{eq:translation_on_field}, gets mapped to the following operation acting on the density matrix $\R$
\[
\label{eq:translation_rho}
\exp(\Delta x L_{\p\x}^+ +\Delta y \frac{L_{\x\x}^+ - L_{\p\p}^+}{2} + \Delta \tau \frac{L_{\x\x}^+ + L_{\p\p}^+}{2})\R =\R'. \]
We can now use Theorem~\ref{theorem:cptp} to impose the CPTP condition on the operation in Eq.~\eqref{eq:translation_rho}. Using the identification in Eq.~\eqref{eq:duality} it reads
\begin{eqnarray}
\Delta \tau^2 &\geq &\Delta x^2+\Delta y^2,\label{eq:cptp_field}\label{eq:causality}\\
   \Delta \tau &\geq& 0.
\end{eqnarray}
The first equation coincides with the causality condition known from special relativity, and the second expresses an assumption about the direction of time. Therefore, we see that the CPTP condition is equivalent to the causality condition, together with the assumption on the direction of time flow.

\begin{figure}[t]
\begin{center}
\includegraphics[width=0.96\hsize]{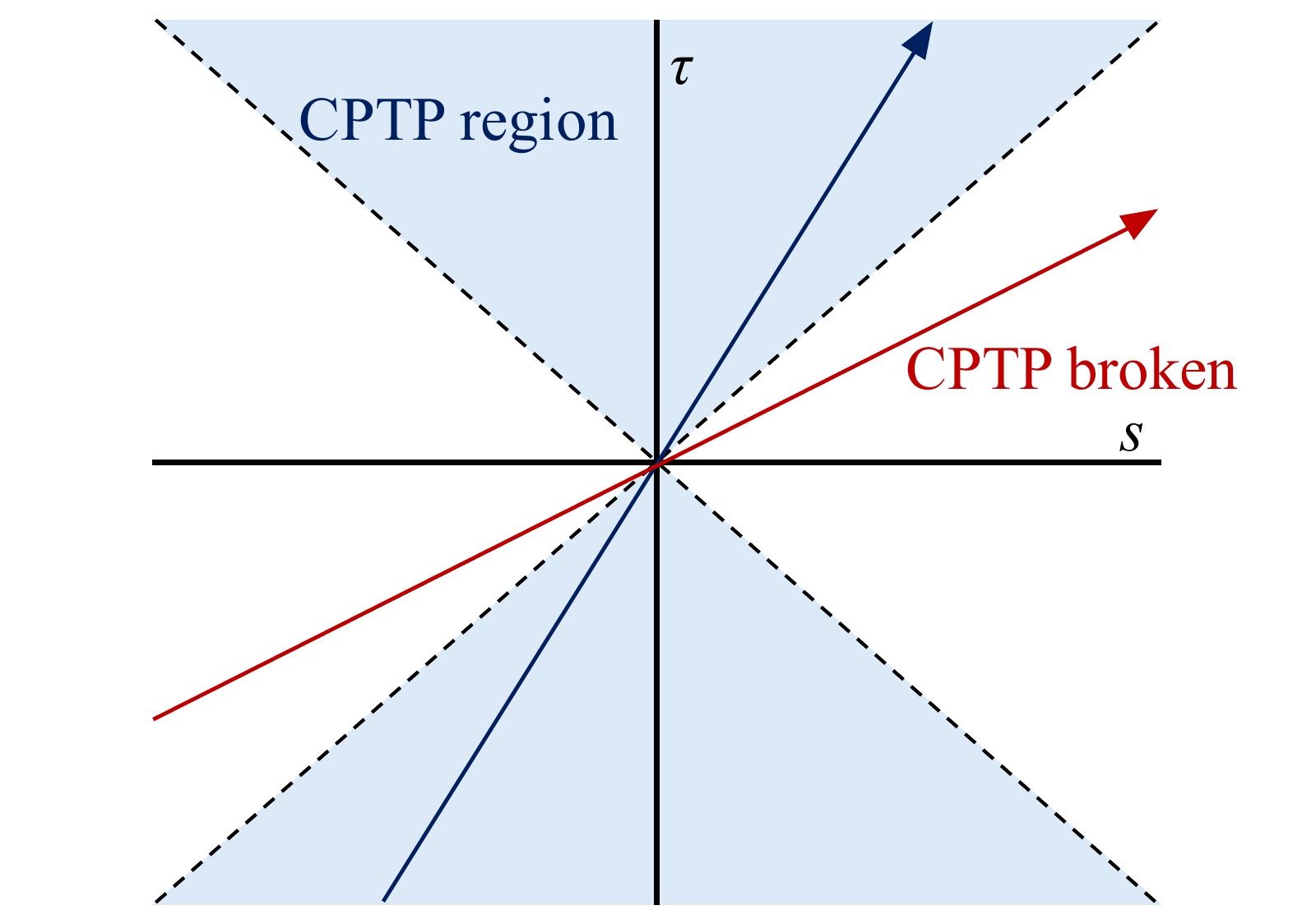}\\
\caption
{CPTP condition on the dual classical field. Time translation has to be equal to or larger than space translation, which means the translation vector lives in the light cone. The blue area inside of the light cone is the CPTP region.
}\label{Fig:lightcon}
\end{center}
\end{figure}

Furthermore, it can be verified that $L_{\x\x}^+L_{\p\p}^+ = (L_{\x\p}^+)^2$ and
the following identity holds for any density matrix $\R$
\[
\label{eq:klein-gordon_open_side}
\left(\frac{(L_{\x\x}^++L_{\p\p}^+)^2}{4}-\frac{(L_{\x\x}^+-L_{\p\p}^+)^2}{4}-(L_{\x\p}^+)^2 \right) \R =0\,.
\]
Using Eq.~\eqref{eq:duality}, we find that this identity is equivalent to the massless Klein-Gordon equation for the field $\phi(x,y,\tau)$
\[
\quad\quad(\partial_t^2-\partial_x^2-\partial_y^2) \phi =0,
\]
since $P_\mu\equiv \partial_\mu$. We recall that in quantum field theory, the Klein-Gordon equation is the equation of motion for a free massless bosonic field. At the same time, the density matrix is an element of the space of operators on a Hilbert space. This suggests a correspondence between the dynamics of a three-dimensional free field theory and the structure of open quantum systems described by density matrices.

Generally, the time $\tau$ which appears as a label of the generator of translation along the time axis $P_\tau$ in the three-dimensional Minkowski space, is not the same as the physical time $t$ that appears in the open quantum dynamics.
However, we find there is some connection.
Let us consider the time translation operator acting on the three-dimensional field $\phi(x,y,\tau)$, \textit{i.e.} $\frac{\partial}{\partial \tau} \phi(x,y,\tau)$. By using Eq.~\eqref{eq:duality}, we get,
\[
\frac{\partial \phi}{\partial \tau} \cong \tfrac{1}{2}(L_{\x\x}^++L_{\p\p}^+)\R %=\frac{d\R}{dt} \,,
\]
The right hand side here represents a valid Lindbladian, therefore, we can consider physical time evolution,
\[\label{eq:align}
\frac{\partial \R}{\partial t} = \tfrac{1}{2}(L_{\x\x}^++L_{\p\p}^+)\R. %=\frac{d\R}{dt} \,,
\]
Thus, the time $\tau$ that labels time translation in the three dimensional Minkowski space can be identified with the physical time $t$ in the open system dynamics with a specific Lindbladian.
In other words, one can relate the time evolution of a field operator satisfying the Klein-Gordon equation in three dimensions to a time evolution of the density matrix from Eq.~\eqref{eq:align}. The dynamics of open quantum systems governed by Eq.~\eqref{eq:align} can be implemented both numerically and experimentally.% 
This means that if we were able to identify an appropriate bijective map $f: \phi \leftrightarrow \R$, we could simulate time evolution of three dimensional massless field, by modeling it as an evolution in the open system dynamics. How exactly to identify this map is an open question that we leave for future work. What we aim to find is a specific map $f$ so that the canonical commutation relations are satisfied. In mathematical terms, let us say we have a field $\phi$ and a corresponding matrix $\R:=f(\phi)$. We solve the open system dynamics for $\R_\tau$, to derive $\phi(x,y,\tau)=f^{-1}(\R_\tau)$. We require that
\[
[\phi(x_1,y_1,\tau), \pi(x_2,y_2,\tau)]=i\delta (x_1-x_2,y_1-y_2),
\]
where $\pi=\partial_\tau\phi$. This is however not guaranteed by all maps $f$. This means that there is no guarantee that the density matrix satisfies the canonical commutation relations. Therefore, establishing a one-to-one correspondence between the field operator and the density matrix is, at this stage, not yet justified.

The correspondence between the two Lie algebras can be further extended to include \textit{all} the $\mathfrak{go}(1)$ generators, going beyond the generators of the Poincar\'e algebra:
The remaining generators in $\mathfrak{go}(1)$ are the displacement operators, $\ad_{\x}$ and $\ad_{\p}$, and the imaginary dissipation operator, $L_{\x\p}^-$. Let us consider the following identifications 
\[\begin{split}\label{eq:duality_super}
    Q_1&\Leftrightarrow -\ad_{\x},\\
    Q_2&\Leftrightarrow \ad_{\p},\\
    D&\Leftrightarrow \tfrac{1}{2}L_{\x\p-},
\end{split}
\]
where $Q_1$, and $Q_2$ can be combined to form a three-dimensional Majorana spinor as follows
\[
Q \equiv \begin{pmatrix}
    Q_1 \\
    Q_2
\end{pmatrix} \Leftrightarrow \begin{pmatrix}
    -\ad_{\x} \\
     \hphantom{-}\ad_{\p}
\end{pmatrix} \, .
\]
$Q_{1,2}$ will be referred to as \textit{supercharges}, while $D$ will be referred to as the \textit{dilation operator}. The rationale for these namings will become clear shortly, as we will demonstrate that these operators form an $\mathcal N = 1$ superconformal algebra in three dimensions.

First, it can be explicitly verified that $-\ad_{\x}$, $\ad_{\p}$ satisfies the following closure relations with the Lorentz operator
\[\begin{split}
\left[-\tfrac{1}{2}\ad_{\N},\begin{pmatrix}
    -\ad_{\x} \\
     \hphantom{-}\ad_{\p}
\end{pmatrix}  \right] &= \begin{pmatrix}
     \hphantom{-}0 & \hphantom{-}\tfrac{1}{2} \\
     -\tfrac{1}{2} & \hphantom{-}0
\end{pmatrix}\begin{pmatrix}
    -\ad_{\x} \\
     \hphantom{-}\ad_{\p}
\end{pmatrix},\\
\left[\hphantom{-}\tfrac{1}{2}\ad_{\hat{X}},\begin{pmatrix}
    -\ad_{\x} \\
     \hphantom{-}\ad_{\p}
\end{pmatrix}  \right] &= \begin{pmatrix}
     \hphantom{-}\tfrac{1}{2} & \hphantom{-}0 \\
     \hphantom{-}0 & -\tfrac{1}{2}
\end{pmatrix}\begin{pmatrix}
    -\ad_{\x} \\
     \hphantom{-}\ad_{\p}
\end{pmatrix},\\
\left[-\tfrac{1}{2}\ad_{\hat{Y}},\begin{pmatrix}
    -\ad_{\x} \\
     \hphantom{-}\ad_{\p}
\end{pmatrix}  \right] &= \begin{pmatrix}
     \hphantom{-}0 & -\tfrac{1}{2} \\
     -\tfrac{1}{2} & \hphantom{-}0
\end{pmatrix}\begin{pmatrix}
    -\ad_{\x} \\
     \hphantom{-}\ad_{\p}
\end{pmatrix},
\end{split}
\]
which corresponds to
 \[
 \label{eq:spinor_def}
[T_{\mu\nu},Q] = -\tfrac{1}{2}\gamma_{\mu\nu} Q \, .
\]
Here, the three-dimensional $\gamma$ matrices are given by $\gamma^\tau=i\sigma^2$, $\gamma^x=\sigma^1$, $\gamma^y=\sigma^3$, $\gamma_{\mu \nu} \equiv \frac{1}{2}(\gamma_\mu \gamma_\nu - \gamma_\nu \gamma_\mu)$, where $\sigma^1,\sigma^2,\sigma^3$ denote $x,y,z$ Pauli matrices, respectively.
Eq.~\eqref{eq:spinor_def} shows that indeed $Q$ is a spinor. Moreover, with these definitions, we can easily verify that the following equation holds,
\[\label{eq:majorana_property}
C\gamma^0=I,
\]
where the charge conjugation matrix is $C=-i\sigma^2$. This is the Majorana property. Finally, $Q_1\cong -\ad_{\x}$ and $Q_2\cong \ad_{\p}$ satisfy the following (anti)-commutation relations
\[\begin{split}
\label{eq:SUSY_algebra}
    \{Q_\alpha,Q_\beta\} = 2(\gamma^\mu C)_{\alpha\beta} P_\mu \, ,
\end{split}
\]
where $P_\mu$ are the translation generators. These anti-commutation relations Eq.~\eqref{eq:SUSY_algebra} indeed form a $\mathcal N = 1$ supersymmetry algebra in three dimensions, thus justifying the name supercharges for $Q_1$ and $Q_2$.

We define the Majorana spinor $\psi$ by the superpartner of the scalar field as,
\[
\psi \equiv 
\begin{pmatrix}
    \psi_1\\
    \psi_2
\end{pmatrix}=
\begin{pmatrix}
    Q_1\\
    Q_2
\end{pmatrix}\phi.
\]
Using the identification between the supercharges and the displacement operator, Eq.~\eqref{eq:duality_super}, we find,
\[
\psi \Leftrightarrow\begin{pmatrix}
    -\ad_{\x}\\
    \hphantom{-}\ad_{\p}
\end{pmatrix}\R.
\]

Similarly to the discussion leading to Eq.~\eqref{eq:klein-gordon_open_side}, the following identity is verified for any density matrix $\R$,
\[\label{eq:Dirac_basic}\begin{split}
\begin{pmatrix}
   \hphantom{-} L_{xp}^+ & \hphantom{-}L_{xx}^+\\
    -L_{pp}^+ & -L_{xp}^+
\end{pmatrix}\begin{pmatrix}
    -\ad_{\x}\\
    \hphantom{-}\ad_{\p}
\end{pmatrix}\R=0,
\end{split}
\]
By applying Eq.~\eqref{eq:duality_super} and $\gamma^\mu$ elements defined above, this identity gets translated into the following conditions for the field $\psi$
\[
\label{eq:Dirac_equation}
\begin{split}
\gamma^\mu \partial_\mu \psi=0.
\end{split}
\]
The first matrix in Eq.~\eqref{eq:Dirac_basic} corresponds to $\gamma^\mu \partial_\mu$. Eq.~\eqref{eq:Dirac_equation} is nothing but the celebrated Dirac equation for a massless spinorial field in three dimensions, which represents the classical equation of motion for a fermionic field operator. 

Furthermore, by identifying the two types of time evolution by $\tau$ and $t$, we conclude that the time evolution of a fermionic field operator obeying the massless Dirac equation can be related to the dynamics of an open quantum system.

In summary, whenever we have both the Klein-Gordon equation and the Dirac equation satisfied, which are in addition connected through the supercharge, this represents the $\mathcal{N} = 1$ supersymmetric free field theory. We have shown exactly this property. Thus, we have demonstrated that the open quantum dynamics for a single bosonic mode maps onto such a theory.
Based on the validity of the Klein-Gordon and Dirac equations, we rephrase this as the following theorem.
\begin{theorem}
    The representation $\mathfrak{go}(1)$ defines a supersymmetric massless free field theory in the three-dimensional Minkowski spacetime. 
\end{theorem}

\section{Superconformal algebra and most general Gaussian superoperator}\label{section:superconformal_algebra}

Up to this point, we have shown that the super-Poincar\'e algebra can be embedded in $\mathfrak{go}(1)$. However, $\mathfrak{go}(1)$ is larger than the super-Poincar\'e algebra, since it also contains the dilation operator, $D\cong \tfrac{1}{2}L_{\x\p}^-$. Let us first show that $D$, indeed, can be interpreted as a dilation operator in three dimensions. It can be verified that the following commutation relations are true
\[\begin{split}\label{eq:D_dilation_properties}
\left[\tfrac{1}{2}L_{\x\p}^-, \begin{pmatrix}
    L_{\x\x}^+\\
    L_{\p\p}^+\\
    L_{\x\p}^+
\end{pmatrix}\right]=\begin{pmatrix}
    L_{\x\x}^+\\
    L_{\p\p}^+\\
    L_{\x\p}^+
\end{pmatrix}\quad
&\Leftrightarrow [D,P_\mu]=P_\mu,\\
\left[\tfrac{1}{2}L_{\x\p}^-, \begin{pmatrix}
    -\ad_{\x}\\
    \hphantom{-}\ad_{\p}
\end{pmatrix}\right]=\frac{1}{2}\begin{pmatrix}
    -\ad_{\x}\\
    \hphantom{-}\ad_{\p}
\end{pmatrix}
&\Leftrightarrow[D,Q_\alpha]=\tfrac{1}{2}Q_\alpha
\end{split}
\]
showing that $D$ is indeed a dilation operator. The presence of the dilation operator, which is an element of a \textit{conformal} algebra, and the validity of both the Dirac and Klein-Gordon equations describing a \textit{massless} field, which is a property of conformal symmetric field theory, suggest a relation with the three-dimensional $N = 1$ \textit{superconformal} algebra, which we demonstrate below. 

Having already shown that $\mathfrak{go}(1)$ is the largest set of trace-preserving superoperators that preserve Gaussianity, it appears that there is no superoperator within $\mathfrak{go}(1)$ corresponding to the special conformal generators, which are the missing ingredients for the appearance of a superconformal algebra. However, Gaussian superoperators corresponding to the special conformal generators can be found. These superoperators lie outside $\mathfrak{go}(1)$. As a consequence, they violate the trace-preserving property. On the other hand, they preserve the signature of the trace distance between different operators, which is the crucial property for the causality structure residing in Eq.~\eqref{eq:causality}.

First, we find the superoperator corresponding to the superconformal charges,
\[\begin{split}
    {S}_1 &\Leftrightarrow -\tfrac{1}{4}\ad^+_{\x} =-\tfrac{1}{2}\{\x,\,\bullet\,\},\\
    {S}_2 &\Leftrightarrow \;\;\,\tfrac{1}{4}\ad^+_{\p} =\;\;\,\tfrac{1}{2}\{\p,\,\bullet\,\}.
\end{split}
\]
Through the Lie super algebra, we can obtain the special conformal operator as
\[\begin{split}
K_\tau &\Leftrightarrow -\tfrac{1}{8}((\ad^+_x)^2+(\ad^+_p)^2),\\
K_x &\Leftrightarrow \tfrac{1}{8}((\ad^+_p)^2-(\ad^+_x)^2),\\
K_y &\Leftrightarrow -\tfrac{1}{4}\ad^+_x\ad^+_p.
\end{split}
\]
Armed with the special conformal operator, $K_\mu$, and superconformal charge, we can define the \emph{extended Gaussian open algebra}, 
\[\begin{split}
\mathfrak{ego}(1) = \mathrm{Span} \big(\ad_{x},\ad_{p},\ad_{\hat{N}},\ad_{\hat{X}},\ad_{\hat{Y}},L_{\x\x}^+,L_{\p\p}^+,L_{\x\p}^+,\\ 
L_{\x\p}^-+\tfrac{1}{2},\ad^+_{\x},\ad^+_{\p},\ad^+_{\x}\ad^+_{\x},\ad^+_{\x}\ad^+_{\p},\ad^+_{\p}\ad^+_{\p} \big).
\end{split}
\]
To close the algebra under commutation relations, we added a constant term to $L_{\x\p}^-$ to redefine the dilation operator as $D\cong \tfrac{1}{2}L_{\x\p}^- +\tfrac{1}{4}$. The redefinition does not affect the commutation relations \eqref{eq:D_dilation_properties}, which made it possible to interpret it as a dilation operator, but it affects other commutation relations. The theorem follows.
\begin{theorem}\label{theorem:superconformal_algebra}
    $\mathfrak{ego}(1)$ is a Lie superalgebra and isomorphic to the $\mathcal{N}=1$ superconformal algebra on three-dimensional spacetime.
\end{theorem}
\begin{proof}
    See Appendix~\ref{section_proof_superconformal}.
\end{proof}

To get $\mathfrak{ego}(1)$, we added five superoperators and redefined one. These changes affected the TP condition, which is no longer satisfied. As a result, a group element generated by an an element of the Lie algebra will no longer be a CPTP map. However, the following theorem says that it will still preserve the Gaussianity. Like the $\mathfrak{go}(1)$ is the largest set of the Gaussianity conserving TP superoperators, we prove that $\mathfrak{ego}(1)$ is the largest set of the Gaussian conserving superoperators. 

\begin{theorem}\label{theorem:iff_condtion_Gaussian_ego}
Let us consider a Gaussian observable, $\hat{O}$, defined as those whose Wigner function is a Gaussian function. Next, consider a linear superoperator, $\mathcal{S}$. If, for any Gaussian observable $\hat{O}$, the time-evolved superoperator $\exp(\mathcal{S}t) \hat{O}$ remains Gaussian for all $t$, then $\mathcal{S}$ belongs to the Lie algebra $\mathfrak{ego}(1)\times \mathrm{Span}(\mathcal{I})$,
where $\mathcal{I}$ denotes the identity superoperator.
\end{theorem}
\begin{proof}
    See Appendix~\ref{section:proof_nogo_ego}.
\end{proof}
We have extended the domain of action of superoperators from the density matrix, $\R$, to Gaussian observables, $\hat{O}$. Since a Gaussian observable is not required to have unit trace, the condition of Gaussian conservation can be defined in a larger space than that of trace-preserving superoperators. In this context, all non-trivial superoperators that preserve Gaussianity are elements of $\mathfrak{ego}(1)$

 The Coleman–Mandula theorem provides the fact that the non-trivial extensions of $\mathbb{R}^3\oplus_{\mathrm{S}}\mathrm{SO}(2,1)$ are extensions only by the super symmetric operator and the conformal operator. Theorem~\ref{theorem:iff_condtion_Gaussian_ego} also provides the fact that the largest extension of the Gaussian conserving superoperators is isomorphic to superconformal algebra. For these reasons, Coleman–Mandula theorem corresponds to Theorem~\ref{theorem:iff_condtion_Gaussian_ego} for the superoperators acting on $\hat{O}$.

Both $\mathfrak{go}(1)$ and $\mathfrak{ego}(1)$ define superoperators on the single bosonic mode. On the other hand, the super Poincar\'e algebra intrinsically includes fermionic degrees of freedom via the supercharge. Therefore,  it is natural to ask how these fermionic degrees of freedom are encoded in the density matrix. To this end, we study how the elements of the density matrix, $\R$, transform under a \textit{three-dimensional} $2\pi$ rotation, $\exp(2\pi T_{xy}) = \exp(-\pi \ad_{\hat{N}})$.

The density matrix after $2\pi$ rotation reads
\[
\R' = \exp(-\pi \ad_{\hat{N}})\R = \exp(-i\pi \an^\dagger \an)\R \exp(i\pi \an^\dagger \an).
\]
Given a certain Fock basis, the density matrix takes the form
\[
\R =\sum_{l,m\geq 0} \R_{lm}\pro{l}{m},
\]
where $\ket{l}$ are number states with integer $l\geq0$. After a $2\pi$ rotation, the elements of $\R$ are evolved to
\[
\R'_{lm} = (-1)^{m-l}\R_{lm}.
\]
Therefore, we see that when $l-m$ is odd, the elements get a negative sign after $2\pi$ rotation. This observation hints towards their fermionic nature under rotations. On the other hand, when $l-m$ is even, there is no negative sign, thus qualifying the corresponding elements as bosonic. In other words, fermionic and bosonic properties under rotations are encoded in the elements of $\R$.
This observation hints at the potential for formulating a Pauli exclusion-like principle, a direction we leave for future work.

\section{Conclusion}

In this work, we have explored all types of open dynamics that preserve the Gaussianity of quantum states. Unlike traditional approaches that rely strictly on the CPTP condition, our method relaxes this requirement to construct a closed Lie algebraic structure. We have named this structure the Gaussian Open Algebra. For an $n$-mode bosonic system, we defined the algebra $\mathfrak{go}(n)$, which is isomorphic to $\mathbb{R}^{2n^2+3n}\oplus_{\mathrm{S}}\mathfrak{gl}(2n,\mathbb{R})$. Furthermore, we demonstrated that the elements of $\mathfrak{go}(n)$ satisfying the CPTP condition correspond to generators of Gaussian quantum channels. 

Building on these results, we solved the time-dependent quadratic order Redfield equation for arbitrary initial states, which can be applied to model systems undergoing decoherence or information loss and studying processes in quantum thermodynamics and transport.

Finally, we showed that within $\mathfrak{go}(1)$, there exists a Lie subalgebra isomorphic to the super Poincaré algebra. We further introduced an extended Lie algebra, called the extended Gaussian open algebra, $\mathfrak{ego}(1)$, which is isomorphic to the $\mathcal{N}=1$ superconformal algebra in three-dimensional spacetime. Within this algebraic structure, we demonstrated that both the massless Dirac equation and the Klein-Gordon equation can be naturally derived, expressed in terms of our representation. Moreover, we discussed the intrinsic relation between the CPTP condition, the principle of causality, and directionality of time, showing how our algebraic framework provides a unified perspective on physical consistency in open quantum dynamics.

These results suggest a fundamental connection between open quantum dynamics and quantum information theory on the one hand, and superconformal field theory on the other. While both areas are major research domains in their own right, their interplay remains largely unexplored. This opens the door to future investigations into whether deeper correspondences exist between central problems in each field, potentially allowing for the translation of challenges from one domain into another, while allowing the sharing of conceptual and technical tools across disciplines.

\section{acknowledgement}
J.-Y.G is supported by the Global-LAMP Program of the National Research Foundation of Korea (NRF) grant funded by the Ministry of Education (No. RS-2023-00301976). J-Y.G would like to thank D. Gang, Y. Baek for the interesting discussions.
DR thanks FAPESP, for the ICTP-SAIFR grant 2021/14335-0 and for the Young Investigator grant 2023/11832-9. DR also acknowledges the Simons Foundation for the Targeted Grant to ICTP-SAIFR. D.\v{S}.~acknowledges the support from the Institute for Basic Science in Korea (IBS-R024-D1). 

\onecolumngrid
\appendix

\section{Proof of Theorem~\ref{theorem:go_is_Lie} and Theorem~\ref{theorem:lie_go(n)}}

\subsection{Single bosonic mode : Theorem~\ref{theorem:go_is_Lie}}\label{subsection:proof_of_iso_go(1)}

$\mathfrak{go}(1)$ is defined as the vector space over the real field, $\mathbb{R}$, spanned by nine basis vectors
\[\begin{split}\label{eq:basis_of_go(1)}
&\ad_{\hat{N}},\;\ad_{\hat{X}},\;\ad_{\hat{Y}},\;\ad_{\hat{x}},\;\ad_{\hat{p}},\\
&L_{\x\x}^{+},\; L_{\x\p}^{+},\; L_{\p\p}^{+},\; L_{\x\p}^{-}.
\end{split}
\]
To show that $\mathfrak{go}(1)$ is Lie algebra, we need to prove the existence of a \emph{bilinear} operator, $[A,B]$, that is closed under $\mathfrak{go}(1)$, and satisfies the \emph{anticommutativity} and the \emph{Jacobi identity}. 
Since the commutation relation of the (super-)operators, $[A,B]=AB-BA$, satisfies the three conditions: bilinear, anticommutativity, and the Jacobi identity, proving closeness is enough to complete the proof.

To prove the closeness and commutation relation between $\mathfrak{go}(1)$, we first rewrite operators in Eq.~\eqref{eq:basis_of_go(1)} in terms of $\ad_{\x}$, $\ad_{\p}$, $\ad^+_{\x}$, and $\ad^+_{\p}$. $\ad_{\hat{O}}^+$ is anti-adjoint operator defined by 
\[
\ad_{\hat{O}}^+ \R = \{\R,\hat{O}\} = \R\hat{O}+\hat{O}\R.
\]
Any operator in Eq.~\eqref{eq:basis_of_go(1)} can be rewritten as a quadratic expression in $\ad_{\x}$, $\ad_{\p}$, $\ad^+_{\x}$, and $\ad^+_{\p}$ as
\[\label{eq:ad_basis_change}\begin{split}
&\ad_{\hat{N}} = \tfrac{1}{2}(\ad_{\x}\ad_{\x}^+ +\ad_{\p}\ad_{\p}^+),\quad \ad_{\hat{X}} = \tfrac{1}{2}(\ad_{\p}\ad_{\x}^+ +\ad_{\x}\ad_{\p}^+),\\
&\ad_{\hat{Y}} = \tfrac{1}{2}(\ad_{\x}\ad_{\x}^+ -\ad_{\p}\ad_{\p}^+),\quad L_{\x\p}^- = \tfrac{1}{2}(\ad_{\x}\ad_{\p}^+ -\ad_{\p}\ad_{\x}^+),\\
&L_{\x\x}^+ = \ad_{\x}\ad_{\x},\quad L_{\x\p}^+ = \ad_{\x}\ad_{\p},\quad L_{\p\p}^+ = \ad_{\p}\ad_{\p},
\end{split}
\]
The reason why we rewrite it this way is to use the simple commutation relation between $\ad_{\x}$, $\ad_{\p}$, $\ad^+_{\x}$, and $\ad^+_{\p}$. There are only two non-trivial commutation relations,
\[\begin{split}
    [\ad_{\x}, \ad_{\p}^+]& =- \ad_{\I}^+=-2,\\
    [\ad_{\p}, \ad_{\x}^+]& = \ad_{\I}^+=2,
\end{split}
\]
while the remaining commutation relations between $\ad_{\x}$, $\ad_{\p}$, $\ad^+_{\x}$, and $\ad^+_{\p}$ are zero. Using this, we can obtain all commutation relations between operators in Eq.~\eqref{eq:basis_of_go(1)}, for example, 
\[
[L_{\x\x}^+, L_{\p\p}^+]=[\ad_{\x}\ad_{\x},\ad_{\p}\ad_{\p}]=0,
\]
by using $[\ad_{\x},\ad_{\p}]=0$.

We classify the operators according to whether they have a zero commutation relation, and make a Lie algebra by spanning them.
\[\begin{split}
    \mathfrak{g}&\equiv\mathrm{Span}(\ad_{\hat{N}},\ad_{\hat{X}},\ad_{\hat{Y}},L_{\x\p}^-)\\
    \mathfrak{h}&\equiv\mathrm{Span}(\ad_{\x},\ad_{\p},L_{\x\x}^+,L_{\p\p}^+,L_{\x\p}^+).
\end{split}
\]
It is obvious $\mathfrak{go}(1)=\mathfrak{g}\times\mathfrak{h}$. Further, $[\mathfrak{h},\mathfrak{h}]=0$, meaning that every element of $\mathfrak{h}$ commutes with every other element of $\mathfrak{h}$, since the commutation relation between $\ad_{\x}$ and $\ad_{\p}$ is zero. What remains is to determine the commutation relations $[\mathfrak{g},\mathfrak{g}]$ and $[\mathfrak{g},\mathfrak{h}]$. Every element $g\in\mathfrak{g}$ can be represented by a corresponding two-by-two real matrix, $\Gamma$,
as 
\[
\begin{pmatrix}
-\ad_{\p} &
\ad_{\x} 
\end{pmatrix} 
\Gamma
\begin{pmatrix}
\frac{\ad^+_{\x}}{2} \\ \frac{\ad_{\p}^+}{2} 
\end{pmatrix} \in \mathfrak{g}.
\]
The basis of $\mathfrak{g}$ is given by
\[\begin{split}\label{eq:gl_2_rep}
\ad_{\hat{N}} &= 
\begin{pmatrix}
-\ad_{\p} &
\ad_{\x} 
\end{pmatrix} 
\begin{pmatrix}
0 & -1 \\
1 & 0 \\
\end{pmatrix}
\begin{pmatrix}
\frac{\ad^+_{\x}}{2} \\ \frac{\ad_{\p}^+}{2} 
\end{pmatrix},\\
\ad_{\hat{X}} &=\begin{pmatrix}
-\ad_{\p} &
\ad_{\x} 
\end{pmatrix} 
\begin{pmatrix}
-1 & 0 \\
0 & 1 \\
\end{pmatrix}
\begin{pmatrix}
\frac{\ad^+_{\x}}{2} \\ \frac{\ad_{\p}^+}{2} 
\end{pmatrix},\;\\
\ad_{\hat{Y}} &= \begin{pmatrix}
-\ad_{\p} &
\ad_{\x} 
\end{pmatrix} 
\begin{pmatrix}
0 & 1 \\
1 & 0 \\
\end{pmatrix}
\begin{pmatrix}
\frac{\ad^+_{\x}}{2} \\ \frac{\ad_{\p}^+}{2} 
\end{pmatrix},\\
L_{\x\p}^{-} &= \begin{pmatrix}
-\ad_{\p} &
\ad_{\x} 
\end{pmatrix} 
\begin{pmatrix}
1 & 0 \\
0 & 1 \\
\end{pmatrix}
\begin{pmatrix}
\frac{\ad^+_{\x}}{2} \\ \frac{\ad_{\p}^+}{2} 
\end{pmatrix}.
\end{split}
\]
With this, using $\tr([\Gamma_1,\Gamma_2])=0$, we can derive the communutation relation relation $[g_1,g_2]$ for any two elements $g_1,g_2\in \mathfrak{g}$ as
\[\begin{split}
[g_1,g_2] &= \left[\begin{pmatrix}
-\ad_{\p} &
\ad_{\x} 
\end{pmatrix} 
\Gamma_1
\begin{pmatrix}
\frac{\ad^+_{\x}}{2} \\ \frac{\ad_{\p}^+}{2} 
\end{pmatrix},\begin{pmatrix}
-\ad_{\p} &
\ad_{\x} 
\end{pmatrix} 
\Gamma_2
\begin{pmatrix}
\frac{\ad^+_{\x}}{2} \\ \frac{\ad_{\p}^+}{2} 
\end{pmatrix}\right] \\
&=\begin{pmatrix}
-\ad_{\p} &
\ad_{\x} 
\end{pmatrix} 
[\Gamma_1,\Gamma_2]
\begin{pmatrix}
\frac{\ad^+_{\x}}{2} \\ \frac{\ad_{\p}^+}{2} 
\end{pmatrix}\in \mathfrak{g}.
\end{split}
\]
Additionally, because $\Gamma_{\hat{N}},\Gamma_{\hat{X}},\Gamma_{\hat{Y}},\Gamma_{L_{\x\p}^-}$ for a basis of two-by-two real matrices, we can rewrite the commutator $[\Gamma_1,\Gamma_2]$ as a linear combination of these matrices. As a result, we can also rewrite $[g_1,g_2]$ as the linear combination of the basis vectors, $\ad_{\hat{N}},\ad_{\hat{X}},\ad_{\hat{Y}},L_{\x\p}^-$, thus, $[g_1,g_2]\in \mathfrak{g}$.

We can also notice that for any two elements, the commutation relation $[g_1,g_2]$ corresponds exactly to the commutation relation of $[\Gamma_1,\Gamma_2]$, meaning that the factors in the resulting linear combination is the same. Hence, these Lie algebras are isomorphic, $\mathfrak{g}\approx\mathfrak{gl}(2,\mathbb{R})$.

Moreover, $\mathfrak{g}$ is a subspace of $\mathfrak{go}(1)=\mathfrak{g}\times\mathfrak{h}$, and due to its closeness, $[\mathfrak{g},\mathfrak{g}]\subset\mathfrak{g}$, $\mathfrak{g}$ forms a Lie algebra. Thus, $\mathfrak{g}$ is a Lie subalgebra of $\mathfrak{go}(1)$.

Finally, we are going to illustrate the remaining commutation relations, $[\mathfrak{g},\mathfrak{h}]$. Each linear element $h_1\in \mathfrak{h}_1\equiv\mathrm{Span}(\ad_{\x},\ad_{\p})$, is represented a two dimensional vector, $\bold{v}$, as
\[\label{eq:ad_vector}
\begin{pmatrix}
    -\ad_{\p} & \ad_{\x} 
\end{pmatrix}\bold{v}=h_1\in \mathfrak{h},
\]
where, the basis is given by
\[
\ad_{\x}=\begin{pmatrix}
    -\ad_{\p} & \ad_{\x} 
\end{pmatrix}\begin{pmatrix}
    0\\
    1
\end{pmatrix},\;\ad_{\p}=\begin{pmatrix}
    -\ad_{\p} & \ad_{\x} 
\end{pmatrix}\begin{pmatrix}
    -1\\
    0
\end{pmatrix}.
\]
With this, we derive the commutation relation between $g\in\mathfrak{g}$ and $h_1\in\mathfrak{h}$,
\[\begin{split}\label{eq:[g,h1]}
[g,h_1] &= \left[\begin{pmatrix}
-\ad_{\p} &
\ad_{\x} 
\end{pmatrix} 
\Gamma
\begin{pmatrix}
\frac{\ad^+_{\x}}{2} \\ \frac{\ad_{\p}^+}{2} 
\end{pmatrix},\begin{pmatrix}
    -\ad_{\p} & \ad_{\x} 
\end{pmatrix}\bold{v}\right] \\
&=\begin{pmatrix}
    \ad_{\p} & -\ad_{\x} 
\end{pmatrix}(\Gamma\bold{v})\in \mathfrak{h},
\end{split}
\]
because $\Gamma\bold{v}$ is also a two-dimensional real vector.

The quadratic ordered operators, $L_{\x\x}^+$, $L_{\p\p}^+$, and $L_{\x\p}^+$, are represented by a real symmetric two-by-two matrix, $D$, as
\[
 \begin{pmatrix}
-\ad_{\p} & \ad_{\x} \\
\end{pmatrix}D\begin{pmatrix}
-\ad_{\p} \\
\ad_{\x} \\
\end{pmatrix}=h_2\in\mathfrak{h}_2,
\]
where the basis of $\mathfrak{h}_2 \equiv\mathrm{Span}(L_{\x\x}^+,L_{\p\p}^+,L_{\x\p}^+)$ is given by
\[\begin{split}
    L_{\x\x}^{+}  &= \begin{pmatrix}
-\ad_{\p} & \ad_{\x} \\
\end{pmatrix}\begin{pmatrix}
0 & 0 \\
0 & 1 \\
\end{pmatrix}\begin{pmatrix}
-\ad_{\p} \\
\ad_{\x} \\
\end{pmatrix},\\
 L_{\p\p}^{+}  &= \begin{pmatrix}
-\ad_{\p} & \ad_{\x} \\
\end{pmatrix}\begin{pmatrix}
1 & 0 \\
0 & 0 \\
\end{pmatrix}\begin{pmatrix}
-\ad_{\p} \\
\ad_{\x} \\
\end{pmatrix},\\
 L_{\x\p}^{+}  &= \begin{pmatrix}
-\ad_{\p} & \ad_{\x} \\
\end{pmatrix}\begin{pmatrix}
0 & -1/2 \\
-1/2 & 0 \\
\end{pmatrix}\begin{pmatrix}
-\ad_{\p} \\
\ad_{\x} \\
\end{pmatrix},
\end{split}
\]

which act on the matrix in Eq.~\eqref{eq:gl_2_rep}. The commutation relation between them with $g\in\mathfrak{g}$ at the matrix representation is given by 
\[\begin{split}\label{eq:[g,h2]}
[g,h_2] &= \left[\begin{pmatrix}
-\ad_{\p} &
\ad_{\x} 
\end{pmatrix} 
\Gamma
\begin{pmatrix}
\frac{\ad^+_{\x}}{2} \\ \frac{\ad_{\p}^+}{2} 
\end{pmatrix},
\begin{pmatrix}
-\ad_{\p} & \ad_{\x} \\
\end{pmatrix}D\begin{pmatrix}
-\ad_{\p} \\
\ad_{\x} \\
\end{pmatrix}\right] \\
&=\begin{pmatrix}
-\ad_{\p} & \ad_{\x} \\
\end{pmatrix}(\Gamma D+D\Gamma^\top)\begin{pmatrix}
-\ad_{\p} \\
\ad_{\x} \\
\end{pmatrix}\in \mathfrak{h},
\end{split}
\]
because $\Gamma D+D\Gamma^\top$ is a real symmetric two-by-two matrix.

Since $\mathfrak{h}=\mathfrak{h}_1\times\mathfrak{h}_2$, we considered all commutation relation between $g\in \mathfrak{g}$ and $h\in \mathfrak{h}$.
As a result, $[\mathfrak{g},\mathfrak{h}] \subset \mathfrak{h}$ means $\mathfrak{h}$ is \emph{ideal} on $\mathfrak{go}(1) = \mathfrak{g}\oplus \mathfrak{h}$.

To summarize the results, $\mathfrak{g}\approx\mathfrak{gl}(2,\mathbb{R})$ is a Lie subalgebra of $\mathfrak{go}(1)$ and $\mathfrak{h}\approx \mathbb{R}^5$ is ideal of $\mathfrak{go}(1)$. The results say that $\mathfrak{go}(1)$ is an extended Lie algebra, $\mathfrak{gl}(2,\mathbb{R})$ by $\mathbb{R}^5$, denoted $\mathfrak{gl}(2,\mathbb{R})\oplus_S\mathbb{R}^5$, where $\oplus_S$ denotes a semidirect sum.

\subsection{Multi bosonic mode: Theorem~\ref{theorem:lie_go(n)}}\label{subsection:proof_of_go(n)}
$\mathfrak{go}(n)$ is the vector space over the real number,$\mathbb{R}$, which has $9n$ basis as
\[\begin{split}\label{eq:single_basis_of_go(n)}
&\ad_{\hat{N_i}},\;\ad_{\hat{X_i}},\;\ad_{\hat{Y_i}},\;\ad_{\hat{x_i}},\;\ad_{\hat{p_i}},\\
&L_{\x_i\x_i}^{+},\; L_{\x_i\p_i}^{+},\; L_{\p_i\p_i}^{+},\; L_{\x_i\p_i}^{-},
\end{split}
\]
for $1\leq i\leq n$ acting on single mode, which we dealt with before. Meanwhile, there is another basis acting on bimode,
\[\begin{split}\label{eq:bi_basis_of_go(n)}
    &\ad_{\hat{N}_{ij+}},\;\ad_{\hat{N}_{ij-}},\;\ad_{\hat{X}_{ij}},\;\ad_{\hat{Y}_{ij}},\\
    &L_{\x_i\x_j}^+,\;L_{\p_i\x_j}^+,\;L_{\x_i\p_j}^+,\;L_{\x_i\x_j}^+,\\
    &L_{\x_i\x_j}^-,\;L_{\p_i\x_j}^-,\;L_{\x_i\p_j}^-,\;L_{\x_i\x_j}^-,\\
\end{split}
\]
for $1\leq i< j \leq n $

We mentioned that the commutation operator $[A, B]$ is a bilinear operator, satisfying the condition for a Lie algebra. To complete the proof, we have to show the closeness of $\mathfrak{go}(n)$ under commutation, and their structure of them.

Like the single mode case, $\mathfrak{go}(1)$, we represent the basis of $\mathfrak{go}(n)$ by using $\ad_{\x_i}$, $\ad_{\p_i}$, $\ad^+_{\x_i}$, and $\ad^+_{\p_i}$.
The single-mode operators at Eq.~\eqref{eq:single_basis_of_go(n)} are equivalent to Eq.~\eqref{eq:basis_of_go(1)}.
\[\begin{split}\label{eq:bimode_for_g}
    &\ad_{\hat{N}_{ij+}} = \tfrac{1}{4}(\ad_{\x_i}\ad_{\x_j}^+ + \ad_{\x_j}\ad_{\x_i}^+ + \ad_{\p_i}\ad^+_{\p_j}+\ad_{\p_j}\ad^+_{\p_i}),\\
    &\ad_{\hat{N}_{ij-}} = \tfrac{1}{4}(\ad_{\p_i}\ad_{\x_j}^+ + \ad_{\x_j}\ad_{\p_i}^+ - \ad_{\x_i}\ad^+_{\p_j} - \ad_{\p_j}\ad^+_{\x_i}),\\
    &\ad_{\hat{X}_{ij}} = \tfrac{1}{4}(\ad_{\x_i}\ad_{\p_j}^+ +\ad_{\p_j}\ad_{\x_i}^+ + \ad_{\x_j}\ad^+_{\p_i}+\ad_{\p_i}\ad^+_{\x_j}),\\
    &\ad_{\hat{Y}_{ij}} = \tfrac{1}{4}(\ad_{\x_i}\ad_{\x_j}^+ +\ad_{\x_j}\ad_{\x_i}^+ - \ad_{\p_i}\ad^+_{\p_j}-\ad_{\p_j}\ad^+_{\p_i}) ,\\
    &L_{\x_i\x_j}^- = \tfrac{1}{2}(\ad_{\x_i}\ad_{\x_j}^+ -\ad_{\x_j}\ad_{\x_i}^+) ,\\
    &L_{\x_i\p_j}^- = \tfrac{1}{2}(\ad_{\x_i}\ad_{\p_j}^+ -\ad_{\p_j}\ad_{\x_i}^+) ,\\    
    &L_{\p_i\x_j}^- = \tfrac{1}{2}(\ad_{\p_i}\ad_{\x_j}^+ -\ad_{\x_j}\ad_{\p_i}^+) ,\\    
    &L_{\p_i\p_j}^- = \tfrac{1}{2}(\ad_{\p_i}\ad_{\p_j}^+ -\ad_{\p_j}\ad_{\p_i}^+) ,\\
\end{split}
\]
and
\[\begin{split}\label{eq:bimode_for_h}
    &L_{\x_i\x_j}^+ =\ad_{\x_i}\ad_{\x_j},\;L_{\x_i\p_j}^+ =\ad_{\x_i}\ad_{\p_j},\;\\
    &L_{\p_i\x_j}^+ =\ad_{\p_i}\ad_{\x_j},\;L_{\p_i\p_j}^+ =\ad_{\p_i}\ad_{\p_j},\;\\
\end{split}
\]

Let us classify the operators
\begin{align}
    \mathfrak{g}&\equiv\mathrm{Span}(\ad_{\hat{N_i}}\!,\ad_{\hat{X_i}}\!,\ad_{\hat{Y_i}}\!,L_{\x_i\p_i}^-\!,\, \mathrm{Eq.~}\eqref{eq:bimode_for_g}),&1\leq i<j\leq n, \label{eq:definition_of_g_in_appen}\\
    \mathfrak{h}_1&\equiv\mathrm{Span}(\ad_{\x_i},\ad_{\p_i}),&1\leq i\leq n,\label{eq:definition_of_h1_in_appen}\\
    \mathfrak{h}_2&\equiv\mathrm{Span}(L_{\x_i\x_i}^+,L_{\p_i\p_i}^+,L_{\x_i\p_i}^+,\, \mathrm{Eq.~}\eqref{eq:bimode_for_h}),&1\leq i<j\leq n.\label{eq:definition_of_h2_in_appen}
\end{align}
We define $\mathfrak{h}\equiv\mathfrak{h}_1\times\mathfrak{h}_2$.
We can write $\mathfrak{go}(n)=\mathfrak{g}\times\mathfrak{h}$ and $[\mathfrak{h},\mathfrak{h}]=0$. 

Let's represent $\mathfrak{g}$ and $\mathfrak{h}$ by finite dimensional matrices. Because we have many adjoint and anti-adjoint operators of $\x_i$ and $\p_i$, it is convenient to use vectorization. 
We define the vector 
\[\begin{split}
    \vec{\ad}=\begin{pmatrix}
    -\ad_{\p_1}\\
    \vdots\\
    -\ad_{\p_n}\\
    \ad_{\x_1}\\
    \vdots\\
    \ad_{\x_n}
\end{pmatrix},\quad
\vec{\ad^+}=\begin{pmatrix}
    \ad_{\x_1}^+\\
    \vdots\\
    \ad_{\x_n}^+\\
    \ad_{\p_1}^+\\
    \vdots\\
    \ad_{\p_n}^+
\end{pmatrix},\quad
\end{split}
\]
For single mode, elements of $\mathfrak{go}(1)$ have one to one relation with a two-by-two real matrix, $\Gamma$. For multi mode, the elements of $\mathfrak{go}(n)$, have one to one relation with a $2n$ by $2n$ real matrix, $\Gamma$, as
\[\frac{1}{2} \vec{\ad}^\top
\Gamma\vec{\ad^+} \in \mathfrak{g}.
\]
The commutation relation $[g_1,g_2]$ for elements of $g_1,g_2\in \mathfrak{g}$ as
 
\[\begin{split}
[g_1,g_2] &= \left[\frac{1}{2} \vec{\ad}^\top
\Gamma_1\vec{\ad^+} ,
\frac{1}{2} \vec{\ad}^\top
\Gamma_2\vec{\ad^+} 
\right] = \tfrac{1}{2} \vec{\ad}^\top[\Gamma_1,\Gamma_2]\vec{\ad^+}\end{split}
\]

The linear ordered operators, $\ad_{\x_i}$, and $\ad_{\p_i}$ are represented by real vector, $\bold{v}$, as
\[
\vec{\ad}^\top\bold{v}=h_1\in\mathfrak{h}_1,
\]
 The commutation relation between them with $g\in\mathfrak{g}$ at the matrix representation is given by 
\[\begin{split}\label{eq:[g,h2]_multi_mode}
[g,h_1] &= \left[
\tfrac{1}{2}\vec{\ad}^\top\Gamma\vec{\ad^+},
\vec{\ad}^\top\bold{v}
\right] =\vec{\ad}^\top
\Gamma \bold{v}\in \mathfrak{h}_1.
\end{split}
\]

The quadratic ordered operator, $L_{\x_i\x_j}^+$, $L_{\p_i\p_j}^+$, and $L_{\x_i\p_j}^+$, are represented by two-by-two symmetric real matrix, $D$, as
\[
\vec{\ad}^\top
D
\vec{\ad}=h_2\in\mathfrak{h}_2,
\]
 The commutation relation between them with $g\in\mathfrak{g}$ at the matrix representation is given by 
\[\begin{split}\label{eq:[g,h2]_multi_mode}
[g,h_2] &= \left[
\tfrac{1}{2}\vec{\ad}^\top\Gamma\vec{\ad^+},
\vec{\ad}^\top
D
\vec{\ad}
\right] =\vec{\ad}^\top
(\Gamma D+D\Gamma^\top)
\vec{\ad}\in \mathfrak{h}_2.
\end{split}
\]

As a result, we know that $[\mathfrak{g},\mathfrak{h}] \subset \mathfrak{h}$, which means $\mathfrak{h}$ is \emph{ideal} on $\mathfrak{go}(1) = \mathfrak{g}\oplus \mathfrak{h}$. The dimension of $\mathfrak{h}$ is $2n^2+3n$.

To summarize the results, $\mathfrak{g}\approx\mathfrak{gl}(2n,\mathbb{R})$ is a Lie subalgebra of $\mathfrak{go}(n)$ and $\mathfrak{h}\approx \mathbb{R}^{2n^2+3n}$ is ideal of $\mathfrak{go}(1)$. The results say that $\mathfrak{go}(1)$ is an extended Lie algebra, $\mathfrak{gl}(2n,\mathbb{R})$ by $\mathbb{R}^{2n^2+3n}$.

\section{Proof of Theorem~\ref{theorem:cptp}}\label{subsection:proof_cptp}

To satisfy the CPTP condition, a Lindblad operator needs to be positive. The positivity of the Lindblad operator indicates that the Lindblad operator has to be express in the following form,
\[\label{eq:general_Lindblad}
\mathcal{L}=\sum_k \hat{A}_k\bullet \hat{A}_k^\dagger-\tfrac{1}{2}\{\hat{A}_k^\dagger \hat{A}_k,\bullet \},
\]
where $A_j$ can be non-Hermitian operators. Let us represent Eq.~\eqref{eq:general_Lindblad} by Hermite operator $\hat{O}_j$ as
\[\label{eq:general_Lindblad_by_hermitian}
\mathcal{L}=\sum_{j,k} \gamma_{jk}\left(\hat{O}_k\bullet \hat{O}_j-\tfrac{1}{2}\{\hat{O}_j \hat{O}_k,\bullet \}\right).
\]
We can define the matrix $\gamma$ composed of $\gamma_{jk}$'s. $\mathcal{L}$ has to be Hermitian, so $\gamma_{kj}$ is complex conjugate of $\gamma_{jk}$, $\gamma_{jk}=\gamma_{kj}^*$. The condition provides the fact that $\gamma$ is hermitian, $\gamma = \gamma^\dagger$.
To make Eq.~\eqref{eq:general_Lindblad_by_hermitian} equivalent to Eq.~\eqref{eq:general_Lindblad}, $\gamma$ has to be a semi-definite positive matrix, $\gamma\geq0$.

At $\mathfrak{go}(1)$, there are four elements related to the Lindblad operator, $L_{xx}^+$, $L_{xp}^+$, $L_{pp}^+$, and $L_{xp}^-$. The linear combination of them is written as
\[\begin{split}\label{eq:lind_by_go}
    \go &= \gamma_{\x\x}^+L_{\x\x}^+ +\gamma_{\x\p}^+L_{\x\p}^+ +\gamma_{\p\p}^+L_{\p\p}^+ +\gamma_{\x\p}^-L_{\x\p}^- \\
    &= 2\gamma_{\x\x}^+\left(\x\bullet\x -\tfrac{1}{2}\{\x^2,\bullet\}\right) +2\gamma_{\p\p}^+\left(\p\bullet\p -\tfrac{1}{2}\{\p^2,\bullet\}\right) \\
    &+(\gamma_{\x\p}^+ +i\gamma_{\x\p}^- )\left(\x\bullet\p -\tfrac{1}{2}\{\p\x,\bullet\}\right) \\
    &+(\gamma_{\x\p}^+ -i\gamma_{\x\p}^- )\left(\p\bullet\x -\tfrac{1}{2}\{\x\p,\bullet\}\right).
\end{split}
\]
To make $\go$ satisfy CPTP condition, $\gamma$ given by,
\[\gamma =
\begin{pmatrix}
    2\gamma_{\x\x}^+ & \gamma_{\x\p}^+ +i\gamma_{\x\p^{-}}\\
    \gamma_{\x\p}^+ -i\gamma_{\x\p^{-}} & 2\gamma_{\p\p}^+
\end{pmatrix},
\]
has to be semi-definite positive. Since $\gamma$ is a two-by-two matrix, it is easy to determine whether $\gamma$ is semi-definite positive or not. $\gamma$ is semi-definite positive, if and only if, $\mathrm{det}(\gamma)\geq 0$, and $\tr(\gamma)\geq$. Hence, $\go$ satisfy the CPTP condition, if and only if, 
\[
\mathrm{det}(\gamma)=4\gamma_{\x\x}^+\gamma_{\p\p}^+-(\gamma_{\x\p}^{+})^2-(\gamma_{\x\p}^{-})^2\geq0,
\]
and 
\[
\tr(\gamma)=\gamma_{\x\x}^++\gamma_{\p\p}^+ \geq 0
\]

\section{Proof of Theorem~\ref{theorem:existence_of_GGO}}\label{section:proof_existence}

We will prove this statement for a multi-mode bosonic system with an arbitrary number of modes, $n$. The result for a single bosonic mode then follows as a special case as $n=1$.

Williamson decomposition allows any Gaussian state to be represented as a Gaussian unitary acting on a tensor product of thermal states, as
\[
\R = \U_G \bigotimes_{i=1}^n \R_{\mathrm{th}}^{(i)}(\beta_i)\U_G^\dagger,
\]
where $\R_{\mathrm{th}}^{(i)}(\beta_i)$ denotes the thermal state of the $i$th mode at inverse temperature $\beta_i$~\cite{Williamson1936, Weedbrook2012}. The thermal state at $\beta$ is 
\[
\R_{\mathrm{th}}(\beta)=(1-e^{-\beta})\sum_{m\geq 0}\pro{m}{m}.
\]
$\mathfrak{go}(n)$ contains all the generators that produce Gaussian unitaries through conjugation. Therefore, there exists an element $\go_{U}$ that satisfies
\[
\R = \exp(\go_{U_G})\bigotimes_{i=1}^n \R_i(\beta_i).
\]
Since unitary transformations are CPTP, $\exp(\go_{U_G})\in \mathrm{GO}^+(n)$ is the CPTP map. Moreover, since the inverse of a unitary operator is also unitary, and thus a CPTP map, it follows that $\exp(-\go_{U_G})\in\mathrm{GO}^+(n)$ is also a CPTP map; Gaussian unitary operators can freely modify the symplectic structure. To prove the proposition, it suffices to show that $\beta_i$ can be adjusted using CPTP elements of $\mathfrak{go}(n)$.

To demonstrate that CPTP elements of $\mathfrak{go}(n)$ can control the temperature, we propose two specific elements, composed of $L_{\x\x}^+$, $L_{\p\p}^+$, and $L_{\x\p}^-$,
\begin{eqnarray}
\go_{\beta_i \downarrow}  &=&\tfrac{1}{4} L_{\x_i\x_i}^+ + \tfrac{1}{4}L_{\p_i\p_i}^+ + \tfrac{1}{2} L_{\x_i\p_i}^-=\an^\dagger_i \bullet \an_i - \tfrac{1}{2}\{\an_i\an_i^\dagger,\bullet\}\\
\go_{\beta_i \uparrow} &=&\tfrac{1}{4} L_{\x_i\x_i}^+ + \tfrac{1}{4}L_{\p_i\p_i}^+ - \tfrac{1}{2} L_{\x_i\p_i}^-=\an_i\bullet\an_i^\dagger -\tfrac{1}{2}\{\an_i^\dagger \an_i,\bullet\}.
\end{eqnarray}
Theorem~\ref{theorem:cptp} guarantees that $go_{\beta_i \downarrow} $ and $\go_{\beta_i \uparrow}$ satisfy the CPTP condition.

From the relation between $\beta_i$ and mean photon number, $\mean{\hat{N_i}}=\tr(\N_i \R_{\mathrm{th}}^{(i)}(\beta_i) )$,
\[\label{eq:relation_beta_N}
\beta_i=\ln(\frac{\mean{\hat{N_i}}+1}{\mean{\hat{N_i}}}),
\]
we can characterize the thermal state by $\mean{\N_i}$. By examining how $\go_{\beta_i \downarrow}$ and $\go_{\beta_i \uparrow}$ transform $\mean{\N_i}$, we can see how it affects $\beta$

Let us first consider the case of $\go_{\beta_i \downarrow}$.
By the adjoint Lindblad equation, we obtain the time derivative of $\hat{N}_i$ by $\go_{\beta_i \downarrow}$ and $\go_{\beta_i\uparrow}$. The adjoint Lindblad equation corresponding to $\go_{\beta_i \downarrow}$ is given by
\[
\frac{d}{dt}\mean{\hat{N}_i}=\mean{\frac{d}{dt}\hat{N}_i}=\mean{\an_i\hat{N}_i\an_i^\dagger -\{\tfrac{1}{2}{\an_i\an_i^\dagger,\hat{N}_i}\}}=\mean{\hat{N}_i}+1,
\]
which follows the 
\[\label{eq:N(t)_down}
\mean{\hat{N}_i}(t)=e^t\left(\mean{\hat{N}_i}(0)+1\right)-1.
\]

Eq.~\eqref{eq:relation_beta_N} and Eq.~\eqref{eq:N(t)_down} yields,
\[
\exp(\go_{\beta_i\downarrow}t)\R_{\mathrm{th}}^{(i)}(\beta_i) = \R_{\mathrm{th}}^{(i)}(\beta_i(t)),
\]
where $\beta_i(t)$ is given by
\[\label{eq:beta_t}
\beta_i(t)=\ln(\frac{e^t\left(\mean{\hat{N}_i}+1\right)}{e^t\left(\mean{\hat{N}_i}+1\right) -1}) \leq \beta_i,
\]
and $\mean{\hat{N}_i} = \frac{1}{e^{\beta_i}-1}$. $\beta_i(t)$ is monotonically decreasing to zero by $t$.

Likewise, we can obtain the adjoint Lindblad equation for $\go_{\beta\uparrow}$, as
\[
\frac{d}{dt}\mean{\hat{N}_i}=\mean{\frac{d}{dt}\hat{N}_i}=\mean{\an_i^\dagger\hat{N}_i\an_i -\{\tfrac{1}{2}{\an_i^\dagger\an_i,\hat{N}_i}\}}=-\mean{\hat{N}_i},
\]
which follows the 
\[
\mean{\hat{N}_i}(t)=e^{-t}\mean{\hat{N}_i}(0),
\]
and we have the relation
\[
\exp(\go_{\beta_i\uparrow}t)\R_{\mathrm{th}}^{(i)}(\beta_i) = \R_{\mathrm{th}}^{(i)}(\beta_i(t)),
\]
where $\beta_i(t)$ is given by
\[\label{eq:beta_t}
\beta_i(t)=\ln(\frac{e^{-t}\mean{\hat{N}_i} +1}{e^{-t}\mean{\hat{N}_i}}),
\]
and $\mean{\hat{N}_i} = \frac{1}{e^{\beta_i}-1}$. $\beta(t)$ is monotonically increasing to infinity by $t$. 

Therefore, we can control $\beta_i$ to take any value in $(0,\infty)$ using $\exp(go_{\beta_i\downarrow}t)$ and $\exp(go_{\beta_i\uparrow}t)$.

Now we are ready to construct the CPTP element $\mathcal{E}_{\R\rightarrow\R'}\in \overline{\mathrm{GO}^+(n)}$
Let's assume that there is two gaussian states, $\R$ and $\R'$, which can be written as
\[
\begin{split}
    \R &= \U_G \bigotimes_{i=1}^n \R_{\mathrm{th}}^{(i)}(\beta_i)\U_G^\dagger,\\
    \R' &= \U_G' \bigotimes_{i=1}^n \R_{\mathrm{th}}^{(i)}(\beta_i')(\U_G')^{\dagger},   
\end{split}
\]
As we discussed, there exist $\exp(-\go_{U_G})\in\mathrm{GO}(n)$ and $\exp(\go_{U'_G})$, can connect $\R$ and $\R'$ to thermal states as
\[
\begin{split}
    \bigotimes_{i=1}^n \R_{\mathrm{th}}^{(i)}(\beta_i)&=\exp(-\go_{U_G})\R,\\
    \R' &= \exp(\go_{U'_G}) \bigotimes_{i=1}^n \R_{\mathrm{th}}^{(i)}(\beta_i'),   
\end{split}
\]

Let's define the set of indices 
$i$ for which $\beta_i\geq\beta_i'$ or $\beta_i<\beta_i'$,
\begin{eqnarray}
    A_{\beta\downarrow}&=&\{1\leq i\leq n | \beta_i\geq\beta_i' \}\\
    A_{\beta\uparrow}&=&\{1\leq i\leq n | \beta_i<\beta_i',\beta_i'\neq \infty \}\\
    A_{\beta=\infty}&=&\{1\leq i\leq n |\beta_i'=\infty \}
    %A_{\beta=\infty}&=&\{1\leq i\leq n | \beta_i'=\infty \}.
\end{eqnarray}
We classified $i$'s not only based on the order between $\beta_i$ and $\beta_i'$, but also on whether $\beta_i'$ is infinity or not. This classification is necessary because when $\beta_i'$ is infinity, there is no element of $\mathrm{GO}^+(n)$ that can connect $\R$ to $\R'$. 
We typically do not consider the case of inverse temperature $\beta=0$, infinite temperature. This is because in that case, the density matrix, $\R$
becomes the identity operator, which is no longer trace-class in an infinite-dimensional Hilbert space. 

If $ A_{\beta=\infty} =\emptyset$, then there exist $t_i\in [0,\infty)$, such that a CPTP element $\mathcal{E}_{\R\rightarrow \R'}\in \mathrm{GO}^+(n)$  can be constructed as
\[
\R' = \mathcal{E}_{\R \rightarrow \R'}\R =\exp(\go_{U_G'}) \prod_{i\in A_{\beta\downarrow}}\exp(\go_{\beta_i\downarrow}t_i) 
\prod_{i\in A_{\beta\uparrow}}\exp(\go_{\beta_i\uparrow}t_i)
\exp(-\go_{U_G})\R
\]

If $ A_{\beta=\infty} \neq\emptyset$, we can instead construct a sequence of CPTP elements $\mathcal{E}_k\in\mathrm{GO}^+(n)$ for $k\in\mathbb{N}$, as
\[
\mathcal{E}_{k} =\exp(\go_{U'}) \prod_{i\in A_{\beta\downarrow}}\exp(\go_{\beta_i\downarrow}t_i) 
\prod_{i\in A_{\beta\uparrow}}\exp(\go_{\beta_i\uparrow}t_i)\prod_{i\in A_{\beta=\infty}}\exp(\go_{\beta_i\uparrow}k)
%\prod_{i\in A_{\beta=\infty}}\exp(\go_{\beta_i\uparrow}k)
\exp(-\go_{U}).
\]
Taking the limit $\lim_{k\rightarrow\infty}\mathcal{E}_k = \mathcal{E}_{\R\rightarrow\R'}\in\overline{\mathrm{GO}^+(n)}$ we obtain a CPTP map in the closure of $\mathrm{GO}^+(n)$ that connects the two Gaussian states,
\[
\R =\mathcal{E}_{\R\rightarrow\R'}\R'
\]
Since we have shown that a CPTP element in $\overline{\mathrm{GO^+(n)}}$ exists for all cases, the statement holds.

\section{Proof of Theorem~\ref{proposition:Gaussian_channel}}\label{section:proof_of_GO+}

We will prove this statement for a multi-mode bosonic system with an arbitrary number of modes, $n$. The result for a single bosonic mode then follows as a special case as $n=1$.

Moreover, for convenience, we will utilize the result of Theorem~\ref{theorem:general_sol_multi} in the proof. It is important to emphasize, however, that the proof of Theorem~\ref{theorem:general_sol_multi} does not depend on Theorem~\ref{proposition:Gaussian_channel}, and thus, the argument in this paper does not involve circular reasoning.

Assume a Gaussian state, $\R$, characterized by an arbitrary covariance matrix $\sigma$ and displacement vector $\bold{d}$, which are defined as
\begin{eqnarray}
    \bold{d}_i &=&\begin{cases} 
		\mean{\x_i} & \mathrm{if}\;i\leq n \\ 
        \mean{\p_{i-n}} & \mathrm{if}\;i>n 
     \end{cases},\\
      \sigma_{ij} &=& \begin{cases} 
		\tfrac{1}{2}\mean{\{\x_i,\x_j\}}-\mean{\x_i}\mean{\x_j} & \mathrm{if}\;i,j\leq n \\ 
       \tfrac{1}{2}\mean{\{\x_i,\p_{j-n}\}}-\mean{\x_i}\mean{\p_{j-n}} & \mathrm{if}\;i\leq n\;\mathrm{and}\;j>n \\
       \tfrac{1}{2}\mean{\{\p_{i-n},\x_{j}\}}-\mean{\p_{i-n}}\mean{\x_{j}}& \mathrm{if}\;i>n\;\mathrm{and}\;j\leq n \\
        \tfrac{1}{2}\mean{\{\p_{i-n},\p_{j-n}\}}-\mean{\p_{i-n}}\mean{\p_{j-n}} & \mathrm{if}\;t>n 
     \end{cases},\\
\end{eqnarray}
where $\mean{\hat{O}}=\tr(\R\hat{O})$.
By using the result of Theorem~\ref{theorem:general_sol_multi},
$\R$ is evolved to $\R'$ under $\mathcal{E}_G=(D\oplus \bold{v})\oplus_{\mathrm{S}}M\in \mathrm{GO}^+(n)$ satisfying CPTP codition.
While there are no constraints on $\bold{v}$ and $D$, the matrix $M$ must satisfy the condition $\mathrm{det}(M)> 0$. $\R'$ is also a Gaussian state with its covariance matrix and displacement vector, as
\begin{eqnarray}
\sigma\; &\overset{\mathcal{E}_G}{\longrightarrow}& \;\sigma' =M\sigma M^\top +2D \label{eq:sigam_tran}\\
\bold{d}\; &\overset{\mathcal{E}_G}{\longrightarrow}& \;\bold{d}'= M\bold{d} + \bold{v}\label{eq:d_tran}.
\end{eqnarray}
where, $\mathrm{det}(M)> 0$, since $M\in \mathrm{GL}^+(2n)$.

Moreover, there exists a sequence $\mathcal{E}_{G,k}\in \mathrm{GO}^+(n)$ such that $\lim_{k\rightarrow\infty}\mathcal{E}_{G,k}\cong (D\oplus \bold{v})\oplus_{\mathrm{S}}M$, where determinant of $M$ is zero. Consequently, $\overline{\mathrm{GO^+(n)}}$ includes elements, $\mathcal{E}_G$ that give rise to transformations, Eq.~\eqref{eq:sigam_tran} and Eq.~\eqref{eq:d_tran},
where $\mathrm{det}(M)\geq0$.

We claim that the CPTP element of $\overline{\mathrm{GO^+}(1)}$ is an infinitesimally divisible Gaussian quantum channel. A Gaussian quantum channel is infinitesimally divisible if and only if:
(1) it transforms the covariance matrix and displacement vector according to Eq.~\eqref{eq:sigam_tran} and Eq.~\eqref{eq:d_tran}, (2)  $\mathrm{det}(M)\geq 0$, and (3) it follows the CPTP condition~\cite{Teiko2009}. We have shown that any CPTP element (condition (3)) of $\overline{\mathrm{GO}^+(n)}$ transforms the covariance matrix and displacement vector as Eq.~\eqref{eq:sigam_tran} and Eq.~\eqref{eq:d_tran} (condition (1)), and that $\mathrm{det}(M)\geq 0$ (condition (2)). Hence, the proposition is proven.

\section{Proof of Theorem~\ref{proposition:Gaussian_channel_2}}\label{section:proof_Gassian=GO}

We will prove this statement for a multi-mode bosonic system with an arbitrary number of modes, $n$. The result for a single bosonic mode then follows as a special case as $n=1$.

As discussed in Section~\ref{section:proof_of_GO+}, we use results of Theorem~\ref{theorem:general_sol_multi} in our proof, but we explicitly clarify that this does not result in circular reasoning. 

We begin by addressing the definition of $\mathrm{GO}(n)$ or a general $n$, which is not included in the main text. To define $\mathrm{GO}(n)=\mathrm{GO}^+(n)\cup\mathrm{GO}^{-}(n)$, we first need to define $\mathrm{GO}^-(n)$, whose definition slightly differs from the case of $n=1$. While $\mathrm{GO}^-(1)$ was defined via the transpose channel, its definition for $\mathrm{GO}^{-1}(n)$ involves the partial transpose channel. 
the partial transpose channel, $\mathcal{E}_{\top_1}$ is defined by
\[
\mathcal{E}_{\top_1}\R = \R^{\top_1},
\]
where $\R^{\top_1}$ denotes the partial transpose on the first mode. For $n=1$ the partial transpose is equivalent to the total transpose, which maintains consistency in the definition.

We are now ready to define $\mathrm{GO}^-(n)$, which is given as
\[
\mathrm{GO}^-(n)=\{\exp(\go)\mathcal{E}_{\top_1}|\go\in\mathfrak{go}(n)\}.
\]
The partial $\mathcal{E}_{\top_1}$ is represented by the matrix,
 \[
 M_{\top_1}=\begin{pmatrix}
     1 & 0 &0_{(2n-2)}\\
     0 & -1&0_{(2n-2)} \\
     0_{(2n-2)} & 0_{(2n-2)} & I_{(2n-2)(2n-2)}
 \end{pmatrix}
 \] 
 with $\mathrm{det}(M)=-1$. By using the multiplicative property of the Lie group, we can see that any element $\mathcal{E}_G\cong (D+\bold{v})\oplus_{\mathrm{S}}M\in \mathrm{GO}^-(n)$ is represented by a matrix $M$ with $\mathrm{det}(M)<0$. 

$\mathrm{GO}(n)$ is union of $\mathrm{GO}^+(n)$ and $\mathrm{GO}^-(n)$, such that $\mathcal{E}_G\cong (D+\bold{v})\oplus_{\mathrm{S}}M \in \mathrm{GO}(n)$ corresponding to $M$ with $\mathrm{det}(M)\neq 0$.
For a state $\R$ characterized by $\sigma$ and $\bold{d}$, $\mathcal{E}_G\cong (D+\bold{v})\oplus_{\mathrm{S}}M\in \mathrm{GO}(n)$ induces the transformation, Eq.~\eqref{eq:sigam_tran} and Eq.~\eqref{eq:d_tran}. Unlike $\mathrm{GO}^+(n)$, $\mathrm{GO}(n)$ includes elements with $\mathrm{det}(M)<0$

Moreover, there exists a sequence $\mathcal{E}_{G,k}\in \mathrm{GO}(n)$ such that $\lim_{k\rightarrow\infty}\mathcal{E}_{G,k}\cong (D\oplus \bold{v})\oplus_{\mathrm{S}}M$, where determinant of $M$ is zero. Consequently, $\overline{\mathrm{GO(n)}}$ includes elements, $\mathcal{E}_G$ that give rise to transformations, Eq.~\eqref{eq:sigam_tran} and Eq.~\eqref{eq:d_tran} with any real matrix $M$.

We claim that the CPTP element of $\overline{\mathrm{GO}(n)}$ is a Gaussian quantum channel. A quantum channel is a Gaussian quantum channel if and only if:
(1) it transforms the covariance matrix and displacement vector according to Eq.~\eqref{eq:sigam_tran} and Eq.~\eqref{eq:d_tran}, and (2) it follows the CPTP condition~\cite{Weedbrook2012}. We have shown that any CPTP element (condition (2)) of $\overline{\mathrm{GO}(1)}$ transforms the covariance matrix and displacement vector as Eq.~\eqref{eq:sigam_tran} and Eq.~\eqref{eq:d_tran} (condition (1)). Hence, the proposition is proven.

\section{Proof of Theorem~\ref{theorem:general_sol_single_mode} and Theorem~\ref{theorem:general_sol_multi}}\label{section:evolution_GO}

We will prove Theorem~\ref{theorem:general_sol_multi}, and the proof of Theorem~\ref{theorem:general_sol_single_mode} follows as a special case with, $n=1$

\subsection{Evolution of Wigner functions under $\mathcal{E}_G$}

First, we consider elements of $\mathrm{GO}^+(n)$, which is $\exp(\go)$, where $\go\in\mathfrak{go}(n)$.
We can decompose $\exp(\go)$ as
\[
\exp(\go) = \exp(h_2)\exp(h_1)\exp(g),
\]
where $g\in\mathfrak{g}$, $h_1\in\mathfrak{h}_1$, and $h_2\in\mathfrak{h}_2$,  The definition of $\mathfrak{h}_1$, $\mathfrak{h}_2$, and $\mathfrak{g}$ is equal to Eq.~\eqref{eq:definition_of_g_in_appen}, Eq.~\eqref{eq:definition_of_h1_in_appen}, and Eq.~\eqref{eq:definition_of_h2_in_appen}.

In section~\ref{subsection:proof_of_go(n)}, we express $g$, $h_1$ and $h_2$ using $\vec{\ad}$ and $\vec{\ad^+}$. As all observables, $\hat{O}$, can be represented through the Wigner function, $W_{\hat{O}}(\vec{x},\vec{p})$, the adjoint and anti-adjoint operators, $\ad$
and $\ad^+$, can likewise be represented as local operators acting on the Wigner function. As operators acting on the Wigner function, $\vec{\ad}$ and $\vec{\ad^+}$ are represented as follows:
\[
\vec{\ad}=-\begin{pmatrix}
    \partial_{x_1}\\
    \vdots\\
    \partial_{x_n}\\
    \partial_{p_1}\\
    \vdots\\
    \partial_{p_n}
\end{pmatrix} =-\begin{pmatrix}
    \vec{\partial_x}\\
    \vec{\partial_p}
\end{pmatrix} \quad \vec{\ad^+}=2\begin{pmatrix}
    {x_1}\\
    \vdots\\
    {x_n}\\
    {p_1}\\
    \vdots\\
    {p_n}
\end{pmatrix}=2\begin{pmatrix}
    \vec{x}\\
    \vec{p}
\end{pmatrix}
\]
Accordingly, $g$, $h_1$ and $h_2$ can be represented as operators acting on the Wigner function as,
\begin{eqnarray}
    g&=&-\begin{pmatrix}
         \vec{\partial_x}^\top &\vec{\partial_p}^\top
    \end{pmatrix}\Gamma \begin{pmatrix}
       \vec{x}\\
        \vec{p}
    \end{pmatrix},\\
    h_1&=&-\begin{pmatrix}
        \vec{\partial_x}^\top &\vec{\partial_p}^\top
    \end{pmatrix} {\bold{v}},\\
    h_2&=&\begin{pmatrix}
        \vec{\partial_x}^\top &\vec{\partial_p}^\top
    \end{pmatrix} D \begin{pmatrix}
        \vec{\partial_x}\\
        \vec{\partial_p}
    \end{pmatrix}.
\end{eqnarray}
As a result, $\exp(\go)$ is given by
\[
\exp(\go) = \exp(\begin{pmatrix}
        \vec{\partial_x}^\top &\vec{\partial_p}^\top
    \end{pmatrix} D \begin{pmatrix}
        \vec{\partial_x}\\
        \vec{\partial_p}
    \end{pmatrix})\exp(-\begin{pmatrix}
        \vec{\partial_x}^\top &\vec{\partial_p}^\top
    \end{pmatrix} {\bold{v}}) \exp(-\begin{pmatrix}
        \vec{\partial_x}^\top &\vec{\partial_p}^\top
    \end{pmatrix}\Gamma \begin{pmatrix}
        \vec{x}\\
        \vec{p}
    \end{pmatrix}),
\]
 is equivalent to the fact that $\exp(\go)$ can be represented in terms of $M=\exp(\Gamma)$, $D$, and $\bold{v}$. Since $M$ is exponential of $\Gamma$,
$\mathrm{det}(M)>0$.
We can derive a closed-form expression for the action of any wingner function $W(\vec{x},\vec{p})$,
\[\label{eq:exp(go)_action}
    W(\vec{x},\vec{p})\overset{\exp(\go)}{\longrightarrow} \frac{1}{\mathrm{det}(M)}\int_{-\infty}^\infty \int_{-\infty}^\infty  d\vec{y} d\vec{q} W\left(M^{-1} 
    \left(\icol{
    \vec{y}\\
    \vec{q}}-\bold{v}\right)
    \right)
    \frac{1}{4\pi\sqrt{\mathrm{det}(D)}}\exp({-\irow{\vec{x}^\top-\vec{y}^\top&\vec{p}^\top-\vec{q}^\top}\tfrac{D^{-1}}{4}\icol{\vec{x}-\vec{y}\\\vec{p}-\vec{q}}}).
\]
In other words, we have obtained a general formula describing how elements in $\mathrm{GO}^+(n)$ transform the Wigner function.

To complete the proof, we also need to consider the elements of $\mathrm{GO}^-(n)$. Any element of $\mathrm{GO}^-(n)$ is expressed by $\exp(\go)\mathcal{E}_{\top_1}$, where $\mathcal{E}_{\top_1}$ is partial transpose channel acting on the first mode. $\mathcal{E}_{\top_1}$ also can be considered as an operator acting on a Wigner function, $W(\vec{x},\vec{p})$,
\[\label{eq:M_top_action}
    W(\vec{x},\vec{p})\overset{\mathcal{E}_{\top_1}}{\longrightarrow} W\left(M_{\top_1}\icol{
    \vec{y}\\
    \vec{q})}\right),
\]
where
 \[
 M_{\top_1}=\begin{pmatrix}
     1 & 0 &0_{(2n-2)}\\
     0 & -1&0_{(2n-2)} \\
     0_{(2n-2)} & 0_{(2n-2)} & I_{(2n-2)(2n-2)}
 \end{pmatrix}.
 \] 
Combining Eq.~\eqref{eq:exp(go)_action} and Eq.~\eqref{eq:M_top_action}, we obtain the how $\exp(\go)\mathcal{E}_{\top_1}$ transforms a Wignerfunction,
\[\label{eq:exp(go)E_top_action}
    W(\vec{x},\vec{p})\overset{\exp(\go)\mathcal{E}_{\top_1}}{\longrightarrow} 
    \frac{1}{|\mathrm{det}(M)|}\int_{-\infty}^\infty \int_{-\infty}^\infty  d\vec{y} d\vec{q} W\left(M^{-1}
    \left(\icol{
    \vec{y}\\
    \vec{q}}-\bold{v}
    \right)\right)
    \frac{1}{4\pi\sqrt{\mathrm{det}(D)}}\exp({-\irow{\vec{x}^\top-\vec{y}^\top&\vec{p}^\top-\vec{q}^\top}\tfrac{D^{-1}}{4}\icol{\vec{x}-\vec{y}\\\vec{p}-\vec{q}}}).
\]
where $M=M_{\top_1}\exp(\Gamma)$. From the facts that $\mathrm{det}(M_{\top_1})=-1$ and $\mathrm{det}(\exp(\Gamma))>0$,  it follows that $\mathrm{det(M)}<0$.

Lastly, we consider the elements of $\overline{\mathrm{GO}(n)}$, i.e., the limits of elements in $\mathrm{GO}(n)$.

We require strong convergence for the limit of any sequence $\mathcal{E}_{G,k}\in \mathrm{GO}(n)$. Strong convergence indicates that $\lim_{k\rightarrow\infty} \mathcal{E}_{G,k} = \mathcal{E}_{G,\infty}$, when 
\[
\lim_{k\rightarrow\infty} \mathcal{E}_{G,k} \R  = \mathcal{E}_{G,\infty} \R,
\]
for all $\R$. Each $\mathcal{E}_{G,k}\cong (D_k\oplus \bold{v}_k)\oplus_{\mathrm{S}}M_k$ consists of a linear coordinate transformation followed by convolution with a Gaussian function. Such operators converge strongly only if the sequences, $M_k$, $\bold{v}_k$, and $D_k$ converge uniformly~\cite{friedlander1999introduction}. 
The sequence $M_k$ can converge to $M_0$ with $\mathrm{det}(M_0)=0$, such that $\overline{\mathrm{GO}(n)}$ include the element $\mathcal{E}_G = (D\oplus \bold{v})\oplus_{\mathrm{S}}M$ where $\mathrm{det}(M)=0$. In the case where $\mathrm{det}(M)=0$, special care must be taken in analyzing the transformation induced by $\mathcal{E}_G$, since the inverse of $M$ is not defined when $\mathrm{det}(M)=0$. To illustrate, we consider the simplest case $\mathcal{E}_{G,0}$ in which $M = 0_{2n\times2n}$, $D = 0_{2n\times2n}$, and $\bold{v} =  0_{2n}$.
\[\label{eq:E_0_action}
    W(\vec{x},\vec{p})\overset{\mathcal{E}_{G,0}}{\longrightarrow} \delta(\vec{x},\vec{p}),
\]
where $\delta(\vec{x},\vec{p})$ is Dirac-delta function.

\subsection{Evolution of $\mathcal{E}_G$}
We will discuss how the elements of $\mathrm{GO}^+(n)$ evolve under $\go'$.
$\mathcal{E}_G\in\mathrm{GO}^+(n)$ can be expressed as $\exp(\go)$ and, as discussed in the previous subsection, can be decomposed into $\exp(h_2)\exp(h_1)\exp(g)$. Each of these can be expressed in terms of $D$, $\bold{v}$, and $M=\exp(\Gamma)$, respectively.
To describe the evolution of $\mathcal{E}_G$, it is sufficient to describe the evolution of $D$, $\bold{v}$, and $M$.

Let us consider a time evolution,
\[\label{eq:dynamics_of_EG}
\frac{d}{dt}\mathcal{E}_G = \go' \mathcal{E}_G,
\]
where we note that $\go$, which serves as the exponent of $\mathcal{E}_G$, and $\go'$ which dictates its evolution.
To distinguish them, we express $\go'$ as
\[
\go'=\frac{1}{2}
\vec{\ad}^\top \Gamma_M \vec{\ad^+}+\vec{\ad}^\top\Gamma_D \vec{\ad}+\vec{\ad}^\top
\Gamma_{\bold{v}}.
\]
Eq.~\eqref{eq:dynamics_of_EG} can be rewritten as
\[\label{eq:dynamics_of_EG_decom}
\begin{split}
    &\frac{d}{dt}\exp(\vec{\ad}^\top D \vec{\ad})\exp(\vec{\ad}^\top
\bold{v})\exp(\frac{1}{2}
\vec{\ad}^\top \Gamma \vec{\ad^+}) \\
&\quad= \left(\frac{1}{2}
\vec{\ad}^\top \Gamma_M \vec{\ad^+}+\vec{\ad}^\top\Gamma_D \vec{\ad}+\vec{\ad}^\top
\Gamma_{\bold{v}}\right) \exp(\vec{\ad}^\top D \vec{\ad})\exp(\vec{\ad}^\top
\bold{v})\exp(\frac{1}{2}
\vec{\ad}^\top \Gamma \vec{\ad^+}).
\end{split}
\]
Applying the commutation relations, the expression can be modified as.
\[\label{eq:dynamics_of_EG_decom}
\begin{split}
    &\frac{d}{dt}\exp(\vec{\ad}^\top D \vec{\ad})\exp(\vec{\ad}^\top
\bold{v})\exp(\frac{1}{2}
\vec{\ad}^\top \Gamma \vec{\ad^+}) \\
    &\quad=\left(\vec{\ad}^\top\frac{dD}{dt}\vec{\ad}\right)
    \exp(\vec{\ad}^\top D \vec{\ad})
    \exp(\vec{\ad}^\top
\bold{v})
\exp(\frac{1}{2} 
\vec{\ad}^\top \Gamma \vec{\ad^+}) \\
&\quad+\exp(\vec{\ad}^\top D \vec{\ad})
\left(\vec{\ad}^\top\frac{d\bold{v}}{dt}\right)
\exp(\vec{\ad}^\top\bold{v})
\exp(\frac{1}{2} \vec{\ad}^\top \Gamma \vec{\ad^+})\\
&\quad+\exp(\vec{\ad}^\top D \vec{\ad})
\exp(\vec{\ad}^\top\bold{v})
\left(\frac{d}{dt}\exp(\frac{1}{2} \vec{\ad}^\top \Gamma \vec{\ad^+})\right)\\
&\quad=\left(\vec{\ad}^\top (\Gamma_D+\Gamma_M D +D\Gamma_M^\top ) \vec{\ad}\right)
    \exp(\vec{\ad}^\top D \vec{\ad})
    \exp(\vec{\ad}^\top
\bold{v})
\exp(\frac{1}{2} 
\vec{\ad}^\top \Gamma \vec{\ad^+}) \\
&\quad+\exp(\vec{\ad}^\top D \vec{\ad})
\left(\vec{\ad}^\top(\Gamma_{\bold{v}}+\Gamma_M\bold{v})\right)
\exp(\vec{\ad}^\top\bold{v})
\exp(\frac{1}{2} \vec{\ad}^\top \Gamma \vec{\ad^+})\\
&\quad+\exp(\vec{\ad}^\top D \vec{\ad})
\exp(\vec{\ad}^\top\bold{v})
\left(\frac{1}{2}\vec{\ad}^\top\Gamma_M\vec{\ad^+}\exp(\frac{1}{2} \vec{\ad}^\top \Gamma \vec{\ad^+})\right).
\end{split}
\]
Comparing the individual terms, we derive three distinct equations,
\begin{eqnarray}
\frac{d}{dt}D &=& \Gamma_D +\Gamma_M D +D\Gamma_M^\top,\\
\frac{d}{dt}\bold{v} &=& \Gamma_{\bold{v}} +\Gamma_M \bold{v},\\
\frac{d}{dt}\exp(\frac{1}{2} \vec{\ad}^\top \Gamma \vec{\ad^+}) &=& \frac{1}{2}\vec{\ad}^\top\Gamma_M\vec{\ad^+}\exp(\frac{1}{2} \vec{\ad}^\top \Gamma \vec{\ad^+}).\label{eq:dynamics_of_M_before}\\
\end{eqnarray}
Using the isomorphism, $\exp(\frac{1}{2} \vec{\ad}^\top \Gamma \vec{\ad^+})$ can be expressed as $M=\exp(\Gamma)$, and Eq.~\eqref{eq:dynamics_of_M_before} is equivalent to,
\[
\frac{d}{dt}M =\Gamma_M M.
\]
Therefore, we have obtained the time evolution of $M$, $D$, and $\bold{v}$.

\section{Proof of theorem~\ref{theorem:superconformal_algebra}}\label{section_proof_superconformal}
$\mathcal{N}=1$ superconformal algebra in three dimensions is isomorphic to $\mathfrak{osp}(1|4)$~\cite{Eberhardt_2021}. Based on this, we will prove the isomorphism between $\mathcal{N}=1$ superconformal algebra in three dimensions and $\mathfrak{ego}(1)$ by establishing the isomorphism between $\mathfrak{ego}(1)$ and $\mathfrak{osp}(1|4)$. 
$\mathfrak{osp}(1|4)$ is composed of ten bosonic generators, $\mathcal{X}_{ij}$ and four fermionic generators, $\mathcal{Y}_{i}$~\cite{Blank1981}, where $i=-2,-1,1,2$. Commutation relations between $\mathcal{X}_{ij}$, and $\mathcal{Y}_{i}$ is given by
\begin{align}
    [\mathcal{X}_{ij},\mathcal{X}_{kl}]&=g_{ik}\mathcal{X}_{jl}+g_{jk}\mathcal{X}_{il}+g_{il}\mathcal{X}_{jk}+g_{jl}\mathcal{X}_{ik},\label{eq:boson_boson_osp}\\
    [\mathcal{X}_{ij},\mathcal{Y}_k]&=g_{ik}\mathcal{Y}_j+g_{jk}\mathcal{Y}_i,\label{eq:boson_fermion_osp}\\
    \{\mathcal{Y}_{i},\mathcal{Y}_j\}&=2\mathcal{X}_{ij}\label{eq:anti_commute_osp},
\end{align}
where $g_{ij}$ is symplectic structure constant as
\[
g_{ij}=\mathrm{sign}(i)\delta_{0,i+j}.
\]

We associate the linear operators, $\ad_{\x}$,  $\ad_{\p}$, $\ad_{\x}^+$, and $\ad_{\p}^+$ with $\mathcal{Y}_i$ as
\[
\mathcal{Y}_2 = -\ad_{\x},\quad\mathcal{Y}_1 = \ad_{\p},\quad\mathcal{Y}_{-1} = \tfrac{1}{2}\ad_{\x}^+,\quad\mathcal{Y}_{-2} = \tfrac{1}{2}\ad_{\p}^+\quad.
\]
Eq.~\eqref{eq:anti_commute_osp} provides the representation of $\mathcal{X}_{ij}$ as the quadratic operators in $\mathfrak{ego}(1)$, explicitly,
\[\begin{split}
\mathcal{X}_{22}&=\tfrac{1}{2}\{\mathcal{Y}_{2},\mathcal{Y}_{2}\}=\ad_{\x}\ad_{\x}=L_{\x\x}^+, \quad
\mathcal{X}_{21}=\tfrac{1}{2}\{\mathcal{Y}_{2},\mathcal{Y}_{1}\}=-\ad_{\x}\ad_{\p}=-L_{\x\p}^+,\\
\mathcal{X}_{11}&=\tfrac{1}{2}\{\mathcal{Y}_{1},\mathcal{Y}_{1}\}=\ad_{\p}\ad_{\p}=L_{\p\p}^+,\quad \mathcal{X}_{-2-2}=\tfrac{1}{2}\{\mathcal{Y}_{-2},\mathcal{Y}_{-2}\}=\tfrac{1}{4}(\ad_{\p}^+)^2,\\
\mathcal{X}_{-2-1}&=\tfrac{1}{2}\{\mathcal{Y}_{-2},\mathcal{Y}_{-1}\}=\tfrac{1}{4}\ad_{\x}^+\ad_{\p}^+,\quad \mathcal{X}_{-1-1}=\tfrac{1}{2}\{\mathcal{Y}_{-1},\mathcal{Y}_{-1}\}=\tfrac{1}{4}(\ad_{\x}^+)^2,\\
\mathcal{X}_{2-1}&=\tfrac{1}{2}\{\mathcal{Y}_{2},\mathcal{Y}_{-1}\}=-\tfrac{1}{4}(\ad_{\x}\ad_{\x}^+ +\ad_{\x}^+\ad_{\x})=-\tfrac{1}{2}\ad_{\hat{N}}-\tfrac{1}{2}\ad_{\hat{Y}}, \\
\mathcal{X}_{2-2}&=\tfrac{1}{2}\{\mathcal{Y}_{2},\mathcal{Y}_{-2}\}=-\tfrac{1}{4}(\ad_{\x}\ad_{\p}^+ +\ad_{\p}^+\ad_{\x})=-\tfrac{1}{2}(L_{\x\p}^- +\tfrac{1}{2})-\tfrac{1}{2}\ad_{\hat{X}},\\
\mathcal{X}_{1-2}&=\tfrac{1}{2}\{\mathcal{Y}_{1},\mathcal{Y}_{-2}\}=\tfrac{1}{4}(\ad_{\p}\ad_{\p}^+ +\ad_{\p}^+\ad_{\p})=\tfrac{1}{2}\ad_{\hat{N}}-\tfrac{1}{2}\ad_{\hat{Y}}, \\
\mathcal{X}_{1-1}&=\tfrac{1}{2}\{\mathcal{Y}_{1},\mathcal{Y}_{-1}\}=\tfrac{1}{4}(\ad_{\p}\ad_{\x}^+ +\ad_{\x}^+\ad_{\p})=-\tfrac{1}{2}(L_{\x\p}^- +\tfrac{1}{2})+\tfrac{1}{2}\ad_{\hat{X}}.
\end{split}
\]
We associated the quadratic operators with $\mathcal{X}_{ij}$, conserving the anti-commutation relation, Eq.~\eqref{eq:anti_commute_osp}, such that, the conservation of commutation relations, Eq.~\eqref{eq:boson_boson_osp} and Eq.~\eqref{eq:boson_fermion_osp}, is sufficient to prove the isomorphism.

In the general $\mathfrak{osp}(1|4)$, the commutation relations among the $\mathcal{Y}_i$ are not defined. However, in our representation, $\mathfrak{ego}(1)$, they satisfy the following commutation relations,
\[
[-\ad_{\x},\tfrac{1}{2}\ad_{\p}^+]=1,\quad[\ad_{\p},\tfrac{1}{2}\ad_{\x}^+]=1,
\]
while the other commutation relations yield zero. Equivalently, this can be written as
\[\label{eq:commute_fermion_osp}
[\mathcal{Y}_i,\mathcal{Y}_j]=g_{ij}.
\]

Accordingly, we have identified the commutation relations, Eq.~\eqref{eq:commute_fermion_osp}, and anti-commutation relations, Eq.~\eqref{eq:anti_commute_osp}, of $\mathcal{Y}_i$ corresponding to elements of  $\mathfrak{ego}(1)$. Moreover, Eq.~\eqref{eq:commute_fermion_osp} and Eq.~\eqref{eq:anti_commute_osp}, guarantee the conservation of the other commutation relations, Eq.~\eqref{eq:boson_boson_osp}, and Eq.~\eqref{eq:boson_fermion_osp}. 

Let us obtain the commutation relation between $\mathcal{X}_{ij}$, using $\mathcal{X}_{ij}=\tfrac{1}{2}(\mathcal{Y}_i\mathcal{Y}_j+\mathcal{Y_j}\mathcal{Y_i})$, and Eq.~\eqref{eq:commute_fermion_osp}, as
\[\begin{split}   
[\mathcal{X}_{ij},\mathcal{X}_{kl}]&=\tfrac{1}{4}[\mathcal{Y}_i\mathcal{Y}_j+\mathcal{Y}_j\mathcal{Y}_i,\mathcal{Y}_k\mathcal{Y}_l+\mathcal{Y}_l\mathcal{Y}_k]\\
&=\tfrac{1}{2}g_{ik}(\mathcal{Y}_j\mathcal{Y}_l + \mathcal{Y}_l\mathcal{Y}_j) 
+\tfrac{1}{2}g_{jk}(\mathcal{Y}_i\mathcal{Y}_l + \mathcal{Y}_l\mathcal{Y}_i)
+\tfrac{1}{2}g_{il}(\mathcal{Y}_j\mathcal{Y}_k + \mathcal{Y}_k\mathcal{Y}_j)
+\tfrac{1}{2}g_{jl}(\mathcal{Y}_i\mathcal{Y}_k + \mathcal{Y}_k\mathcal{Y}_i)\\
&=g_{ik}\mathcal{X}_{jl}+g_{jk}\mathcal{X}_{il}+g_{il}\mathcal{X}_{jk}+g_{jl}\mathcal{X}_{ik},
\end{split}
\]
which corresponding to Eq.~\eqref{eq:boson_boson_osp}. 

Likewise, Let us obtain the commutation relation between $\mathcal{X}_{ij}$ and $\mathcal{Y}_k$,
\[\begin{split}
[\mathcal{X}_{ij},\mathcal{Y}_{k}]= \tfrac{1}{2}[\mathcal{Y}_i\mathcal{Y}_j+\mathcal{Y}_j\mathcal{Y}_i,\mathcal{Y}_k]=g_{ik}\mathcal{Y}_j + g_{jk}\mathcal{Y}_i
\end{split}
\]
which corresponding to Eq.~\eqref{eq:boson_fermion_osp}. Consequently, we have recovered Eq.~\eqref{eq:boson_boson_osp}, and Eq.~\eqref{eq:boson_fermion_osp} by using the commutation and anti-commutation relations of $\mathcal{Y}_i$.

Establishing a one-to-one correspondence between the elements of $\mathfrak{ego}(1)$ and $\mathfrak{osp}(1|4)$, and demonstrating the preservation of all (anti-)commutation relations, we have proven that the two Lie superalgebras are isomorphic.
\section{Proof of theorem~\ref{theorem:iff_condtion_Gaussian_ego}}\label{section:proof_nogo_ego}

We will use proof by contradiction. For contradiction, let us assume that there exists a linear operator $\mathcal{S}$ which conserves the Gaussianity, which does not belong into $\mathfrak{ego}(1)\times\mathrm{Span}(\mathcal{I})$, $\mathcal{S}\notin \mathfrak{ego}(1)\times\mathrm{Span}(\mathcal{I})$. We will show that $\mathcal{S}$ must belong to $\mathfrak{ego}(1)\times\mathrm{Span}(\mathcal{I})$, otherwise there is a Gaussian observable $\hat{O}$ losing its Gaussianity, which is a contradiction.

Since $\mathcal{S}$ is a linear operator, we can express $\mathcal{S}$ as
\[
\mathcal{S} = \sum_j  \hat{A}_j\bullet \hat{B}_j.
\]
Any operator, $A_k$ and $B_k$ can be decomposed by $\an$ and $\an^\dagger$ as
\[\begin{split}
    \hat{A}_j = \sum_{k,l} A_{j,kl}  \an^{\dagger k}\an^{l},\\
    \hat{B}_j = \sum_{k,l} B_{j,kl}  \an^{\dagger k}\an^{l},
\end{split}
\]
for non-negative integer, $k$ and $l$. This is because monomials $\an^{\dagger k}\an^{l}$ form an operator basis in a single-mode bosonic space~\cite{Cahill1969}.

Likewise, we can write the $\mathcal{S}$ using a series expansion as
\[\label{eq:properordered_LG}
\mathcal{S}=\sum_{j,k,l,m} \gamma_{jklm} \an^{\dagger j}\an^{ k}\bullet \an^{ l}\an^{\dagger m}.
\]
with complex coefficients $\gamma_{jklm}$, where $j,k,l$ and $m$ are non-negative integers.
$\an$ and $\an^\dagger$ is a linear combination of $\x$ and $\p$, $\mathcal{S}$ can be rewritten by
\[\label{eq:properordered_LG_xp}
\mathcal{S}=\sum_{j,k,l,m} \gamma_{jklm}' \x^j\p^{ k}\bullet \p^{ l}\x^{m}.
\]
$\x$ acting on $\hat{O}$ from the left side is equivalent to $\tfrac{1}{2}(\ad_{\x}^+ -i \ad_{\x})$, i.e., $\x\hat{O}=\tfrac{1}{2}(\ad_{\x}^+(\hat{O}) -i \ad_{\x}(\hat{O}))$, $\x$ acting from the right is equivalent to $\tfrac{1}{2}(\ad_{\x}^+ +i \ad_{\x})$. The action of $\p$ on the left and right sides is expressed similarly. Hence, Eq.~\eqref{eq:properordered_LG_xp} is rewritten as
\[\begin{split}\label{eq:properordered_LG_ad}
    \mathcal{S}&=\sum_{j,k,l,m} \frac{\gamma_{jklm}'}{2^{j+k+l+m}}  
    (\ad_{\hat{x}}^+ +i \ad_{\hat{x}})^m
    (\ad_{\hat{p}}^+ +i \ad_{\hat{p}})^l
    (\ad_{\hat{x}}^+ -i \ad_{\hat{x}})^j
    (\ad_{\hat{p}}^+ -i \ad_{\hat{p}})^k\\
    &=\sum_{j,k,l,m} \alpha_{jklm} (\ad_{\x}^+)^j(\ad_{\p}^+)^k(\ad_{\x})^l(\ad_{\p})^m.
\end{split}
\]
Since it appears in $\frac{d\hat{O}}{dt}=\mathcal{S}\hat{O}$, to preserve Hermitianity of $\hat{O}(t)=\exp(\mathcal{S}t)\hat{O}$, we require $\mathcal{S}\hat{O}$ to be a Hermitian operator. Additionally, note that for every combination of powers, $(\ad_{\x}^+)^j(\ad_{\p}^+)^k(\ad_{\x})^l(\ad_{\p})^m\hat{O}$ is a Hermitian operator. This follows from combining a) for any Hermitian operator $\hat{H}$, also $\ad_{\hat{H}}\hat{O}$, and $\ad^+_{\hat{H}}\hat{O}$ are Hermitian, thus also b) $(\ad_{\hat{H}_1}\ad_{\hat{H}_2}\hat{O} )^\dagger=\ad_{\hat{H}_1}\ad_{\hat{H}_2}\hat{O}$, and we can use induction to prove the statement for every combination of powers. This implies that all $\alpha_{jklm}$ are real. Thus, we have shown that $\mathcal{S}$ must necessarily have the form of Eq.~\eqref{eq:properordered_LG_ad} with all $\alpha_{jklm}$ real.

Next, we consider the property of a Gaussian-conserving operator. We consider Gaussian observables which have a Gaussian function as the Wigner function as,
\[
W(x,p) = A\exp({-\irow{x-x_0&p-p_0}\Lambda\icol{x-x_0\\p-p_0}}).
\]
If a Gaussian observable, $\hat{O}$, is evolved to another Gaussian observable by $\mathcal{S}$, the Wigner function must preserve its form. It follows that we can express the evolution via the chain rule as
\[\label{eq:L_GR}
\mathcal{S}\hat{O}(t) = \frac{d}{dt}\hat{O}(t) =\left(\lambda_A\frac{\partial}{\partial A}+\lambda_\Lambda\frac{\partial}{\partial \Lambda}+\lambda_{x_0}\frac{\partial}{\partial x_0} + \lambda_{p_0}\frac{\partial}{\partial p_0}\ \right)W(x,p),
\]
where $\lambda_A =\frac{\partial A}{\partial t}$, $\lambda_\Lambda=\frac{\partial \Lambda}{\partial t}$, $\lambda_{x_0}=\frac{\partial x_0}{\partial t}$, and $\lambda_{p_0}=\frac{\partial {p_0}}{\partial t}$.
Executing the derivatives we find that Eq.~\eqref{eq:L_GR} can be rewritten as
\[\label{eq:L_GR_2}
L_G \R =(\lambda_{x^2}x^2+\lambda_{p^2}p^2+\lambda_{xp}xp+\lambda_{x}x+\lambda_{p}p+\lambda_{0} )W(x,p),
\]
where $\lambda$'s are real coefficients. This means that for a Gaussian state, for an operator $L_G$ to preserve its Gaussianity, the Wigner function being acted upon by $L_G$ must transform as being multiplied by at most quadratic powers in $x$ and $p$. 
%Eq.~\eqref{eq:L_GR_2} indicates that $L_G\R$ has Wigner function as a product of $\R$ with up to quadratic order of $x$ and $p$. Moreover, since we require that $L_G$ conserve the Gaussianity for any Gaussian state, $L_G\R$ has to be express as Eq.~\eqref{eq:L_GR_2} for any Gaussian state.

Now we check how the $L_G$, which must have a form of~\eqref{eq:properordered_LG_ad} with all $\alpha_{jklm}$ real, transforms the Wigner function, and compare it to the requirement given by Eq.~\eqref{eq:L_GR_2}.

%We need to check that $L_G$ at Eq.~\eqref{eq:properordered_LG_ad} generate a Wigner function of $L_G\R$ with a Gaussian state as Eq.~\eqref{eq:L_GR_2}. 

The adjoint operators acting on the density matrix are transformed into the operators acting on the Wigner function as~\cite{Breuer2002}
\[\begin{split}
\ad_{\x} \rightarrow- \frac{\partial}{\partial p},\quad
\ad_{\p} \rightarrow \frac{\partial}{\partial x},\quad
\ad_{\x}^{+}  \rightarrow 2x,\quad
\ad_{\p}^{+}  \rightarrow 2p,
\end{split}
\]
which provides
\[
    \mathcal{S}=\sum_{j,k,l,m} \alpha_{jklm} (2x)^j(2p)^k\left(-\frac{\partial}{\partial p}\right)^l\left(\frac{\partial}{\partial x}\right)^m.
\]
Consider a simple Gaussian observable, $\hat{O}$, with the Wigner function
\[
W(x,p)=\exp(-\Lambda_{xx}x^2 -\Lambda_{pp}p^2 ),
\]
where $\Lambda_{xx}>0$ and $\Lambda_{pp}>0$
Applying $\mathcal{S}$ (recall Eq.~\eqref{eq:properordered_LG_ad}) to this Wigner function gives,
\[\begin{split}\label{eq:L_GR_poly}
    \mathcal{S}\hat{O} &=\!\! \sum_{j,k,l,m} \!\alpha_{jklm} (2x)^j(2p)^k (\sqrt{\Lambda_{pp}})^l (-\sqrt{\Lambda_{xx}})^m H_l(\sqrt{\Lambda_{pp}}p)H_m(\sqrt{\Lambda_{xx}}x) \exp(-\Lambda_{xx}x^2 -\Lambda_{pp}p^2 ),
\end{split}
\]
where $H_n(x)$ are Hermite polynomials. Using the following equation,
\[
x^j = \frac{i!}{2^j}\sum_{i\geq 0}^{\floor{j}}\frac{1}{i!(j-2i)!}H_i(x),
\]
which holds for any non-negative integers, $i$ and $j$,
we decompose Eq.~\eqref{eq:L_GR_poly} as
\[\label{eq:LGonR}
\begin{split}
    \mathcal{S}\hat{O} &=\!\! \sum_{j,k,l,m\geq 0} \!\alpha'_{jklm} (2x)^j(2p)^k (\sqrt{\Lambda_{pp}})^l (-\sqrt{\Lambda_{xx}})^m 
    \left(\frac{l!}{2^l}\sum_{i}^{\floor{\tfrac{l}{2}}}\frac{1}{i!(l-2i)!}H_i(\sqrt{\Lambda_{pp}}p) \right)
    \left(\frac{m!}{2^m}\sum_{i}^{\floor{\tfrac{m}{2}}}\frac{1}{j!(m-2i)!}H_j(\sqrt{\Lambda_{xx}}x) \right)\\
    &\times\exp(-\Lambda_{xx}x^2 -\Lambda_{pp}p^2 )\\
    &= \sum_{j,k,l,m\geq 0} \!\alpha'_{jklm} (2x)^j(2p)^k (\Lambda_{pp}p)^l (-\Lambda_{xx}x)^m \exp(-\tfrac{1}{4D_{xx}}x^2 -\tfrac{1}{4D_{pp}}p^2 ),
\end{split}
\]
where $\alpha'_{knlm}$'s are linear combination of $\alpha_{knlm}$. We compare Eq.~\eqref{eq:LGonR} with Eq.~\eqref{eq:L_GR_2}. Since in our example we required $\Lambda_{xx}>0$ and $\Lambda_{pp}>0$, to satisfy the Gaussianity preserving condition, we have  $\alpha'_{klmn}=0$ for any $k+l+m+n>2$. 

Considering linear relation between $\alpha_{knlm}$ and $\alpha_{knlm}'$, we can notice that $\alpha_{knlm}$ also have to be zero whenever $k+n+l+m>2$. This means that $L_G$ has to be expressed by
\[\begin{split}\label{eq:properordered_LG_ad_2}
    \mathcal{S}
    &=\sum_{j\geq 0,k\geq 0,l\geq 0,m\geq 0,j+k+l+m\leq 2} \alpha_{jklm} (\ad_{\x}^+)^j(\ad_{\p}^+)^k(\ad_{\x})^l(\ad_{\p})^m
\end{split}
\]

As a result, $\mathcal{S}$ is composed of adjoint and anti-adjoint operators of $\x$ and $\p$, up to quadratic order.

$\mathfrak{ego}(1) \times \mathrm{Span}(\mathcal{I})$ contains all linear superoperators formed from adjoint and anti-adjoint operators of $\x$ and $\p$ up to quadratic order. Hence, $\mathcal{S}$ must belong to $\mathfrak{ego}(1) \times \mathrm{Span}(\mathcal{I})$, which contradicts the assumption.
\twocolumngrid

\bibliography{bib.bib}
\end{document}